\newcommand{\beq}{\begin{equation}}
\newcommand{\eeq}{\end{equation}}
\newcommand{\bea}{\begin{eqnarray}}
\newcommand{\eea}{\end{eqnarray}}

\newcommand{\gsim}{\lower.7ex\hbox{$\;\stackrel{\textstyle>}{\sim}\;$}}
\newcommand{\lsim}{\lower.7ex\hbox{$\;\stackrel{\textstyle<}{\sim}\;$}}

\newcommand{\trh}{T_{\rm RH}}
\newcommand{\arh}{a_{\rm RH}}

\documentclass[a4paper,11pt]{article}
\pdfoutput=1

\usepackage{jcappub}
\usepackage[utf8]{inputenc} 
\usepackage{amsfonts,amssymb,mathrsfs,amsmath,esint,bm}
\usepackage{pdflscape} 
\usepackage[capitalise]{cleveref}
\allowdisplaybreaks

\graphicspath{{./Figures/}}
\usepackage{latexsym}
\usepackage{graphicx}
\usepackage{subfigure}
\usepackage[dvipsnames]{xcolor}
\usepackage{booktabs}
\usepackage{physics}
\usepackage{tikz}
\usetikzlibrary{decorations.pathmorphing}\usetikzlibrary{positioning,calc}
\usetikzlibrary{math}
\usepackage{cleveref}
\usepackage{comment}
\usepackage[normalem]{ulem}

\hypersetup{pdfstartview=FitV,colorlinks=true,linkcolor=blue,citecolor=red,filecolor=black,urlcolor=blue}

\def\stacksymbols #1#2#3#4{\def\theguybelow{#2}
    \def\vp{\lower#3pt}
    \def\sp{\baselineskip0pt\lineskip#4pt}
    \mathrel{\mathpalette\intermediary#1}}

\def\intermediary#1#2{\vp\vbox{\sp
     \everycr={}\tabskip0pt
     \halign{$\mathsurround0pt#1\hfil##\hfil$\crcr#2\crcr
              \theguybelow\crcr}}}

\title{Reheating in No-Scale Models of Inflation}

\author[a,b]{\bf Ignatios Antoniadis,}
\affiliation[a]{School of Natural Sciences, Institute for Advanced Study, Princeton, NJ 08540, USA}
\affiliation[b]{Laboratoire de Physique Th\'eorique et Hautes \'Energies
  - LPTHE,
Sorbonne Universit\'e, CNRS, 4 Place Jussieu, 75005 Paris, France
}
\emailAdd{antoniad@lpthe.jussieu.fr}

\author[c,d]{John Ellis,}
\affiliation[c]{Theoretical Particle Physics and Cosmology Group, Department of
  Physics, \\ King's~College~London, London WC2R 2LS, United Kingdom}
\affiliation[d]{Theoretical Physics Department, CERN, CH-1211 Geneva 23,
  Switzerland}
\emailAdd{john.ellis@cern.ch}

\author[e,f]{\bf \\ Dimitri V. Nanopoulos}
\affiliation[e]{Academy of Athens, Division of Natural Sciences, Athens 10679, Greece
}
\affiliation[f]{George P. and Cynthia W. Mitchell Institute for Fundamental Physics and Astronomy, Texas A\&M University, College Station, TX 77843, USA
}
\emailAdd{dimitri@physics.tamu.edu}

\author[g]{Keith A. Olive,}
\affiliation[g]{William I.~Fine Theoretical Physics Institute, School of Physics and Astronomy, University of Minnesota, Minneapolis, MN 55455, USA}
\emailAdd{olive@umn.edu}

\author[h]{and Sarunas~Verner,}
\affiliation[h]{Kavli Institute for Cosmological Physics,
University of Chicago, 5640 South Ellis Ave., Chicago, IL 60637, USA}
\emailAdd{verner@uchicago.edu}

\abstract{Analogously to the suppression of inflaton decays into conformally-coupled scalar fields in the original Starobinsky $R + R^2$ model of inflation, inflaton decays to Standard Model fields are also suppressed in minimal no-scale models of inflation with field space curvature $\mathcal{R} = 2/3$. We study how this suppression can be avoided in generalized no-scale inflationary models. These include models in which the field space curvature $\mathcal{R} = 2/(3\alpha)$ with $\alpha \ne 1$ as exemplified by models derived from string theory, as well as models with non-minimal gauge kinetic terms and anomaly-induced couplings. We analyze direct and anomaly-induced inflaton couplings to gauge bosons and gauginos and demonstrate the K\"ahler-frame invariance of the physical gauge coupling. We determine the resulting reheating temperatures and the corresponding predictions in the $(n_s,r)$ plane. Finally, we consider an $R^3$ deformation of Starobinsky supergravity, which modifies the inflaton and stabilizer sectors but does not, by itself, generate new tree-level inflaton couplings to visible matter fields.}

\begin{document}
\begin{flushright}
UMN--TH--4533/26, FTPI--MINN-26/13,   \\
KCL--PH--TH/2026-19, CERN--TH--2026-154 \\
June  2026
\end{flushright}
\maketitle

\section{Introduction}

One of the most promising models of inflation is that proposed originally by Starobinsky, which was derived from an $R+R^2$ theory of gravity~\cite{Staro}. When transformed to the Einstein frame, the theory is described by Einstein gravity with a single real scalar field, $\phi$, with potential \cite{WhittStelle,Kalara:1990ar}
\beq
V(\phi) = \frac{3}{4} m^2  M_{P}^{2}\left(1-e^{-\sqrt{\frac{2}{3}} \frac{\phi}{M_{P}}}\right)^{2} \, ,
\label{Staro}
\eeq
where $M_P \simeq 2.4 \times 10^{18}$~GeV is the reduced Planck mass. Inflation (near-exponential expansion) occurs at large values of $\phi$, where the potential is almost flat. This expansion 
ceases at $\phi_{\rm end}\simeq 0.63 M_P$ when the field enters into harmonic oscillations about its minimum at $\phi = 0$. The inflaton mass for this model is simply $m$. 

While an extended period of exponential expansion solves a host of cosmological problems \cite{Guth:1980zm} and provides a source for cosmological perturbations \cite{MC,pert}, the return to a standard radiation-dominated phase in the early universe is an essential feature of any inflation model \cite{reviews}. So long as the inflaton is coupled (in some way) to Standard Model (SM) fields, reheating is possible
during the oscillatory phase after inflation. How this coupling arises in models of inflation based on no-scale supergravity \cite{no-scale} is the focus of this paper. In particular, we are interested in no-scale models which are capable of reproducing Starobinsky(-like) models of inflation \cite{eno6,building}. 

In the Starobinsky model \cite{Staro}, originally formulated as an $R+R^2$ theory of gravity, one may introduce
the SM action in the Jordan frame with a minimal coupling to curvature. In this case,  
one naturally finds a coupling of the inflaton to the Higgs boson in the Einstein frame \cite{Ema:2024sit}, leading to the possibility of inflaton decay with a rate 
\begin{align}
	\Gamma(\phi \to HH) = \frac{N_H}{192\pi}\frac{m^3}{M_P^2}\,,
	\label{eq:decay_R2_A1}
\end{align}
where $N_H = 4$ is
the number of the real scalar degrees of freedom of the SM Higgs doublet. 
Inflaton decays quickly lead to a thermal bath that redshifts as $\rho_{\rm R} \propto a^{-3/2}$ \cite{Scherrer:1984fd,Giudice:2000ex,GKMO1,GKMO2} where $a$ is the cosmological scale factor. This can be compared to the (initially dominant) inflaton energy density $\rho_\phi$ that redshifts as $a^{-3}$. The reheating temperature is defined by equality between the inflaton and thermal energy densities: $\rho_\phi (\arh) = \rho_{\rm R}(\arh)$, i.e.,
\beq
\frac{g_*(\trh) \pi^2}{30} \trh^4 = \frac{12}{25} \Gamma_\phi^2 M_P^2 \, ,
\label{TRH}
\eeq
where $g_*(\trh)$ is the number of relativistic degrees of freedom at $\trh$.
The di-Higgs decay rate Eq.~(\ref{eq:decay_R2_A1})  is important, as it leads to natural reheating with a reheating temperature 
\beq
\trh =  2.9 \times 10^9~{\rm GeV} \left(\frac{m}{3 \times 10^{13}~{\rm GeV}} \right)^\frac32 \, ,
\eeq
where the inflaton mass $m= 3\times 10^{13}$~GeV is fixed by the amplitude of the scalar perturbation spectrum \cite{Planck} and we have taken $g_*(\trh) = 427/4$.~\footnote{We note that later, when we consider a supersymmetric framework, we use the minimal supersymmetric Standard Model (MSSM) value of $g_*(\trh) = 915/4$.}

However, if the Higgs boson has a coupling $\xi$ to curvature in the Jordan frame, the rate is modified so that \cite{Watanabe:2006ku,Ema:2024sit}
\begin{align}
	\Gamma(\phi \to HH) = \frac{N_H(1-6\xi)^2}{192\pi}\frac{m^3}{M_P^2}\,,\label{ratea2ein}
\end{align}
and the rate vanishes for the conformal coupling $\xi = 1/6$. 

However, even if SM fields are conformally coupled to curvature, the possibility of inflaton decay persists in the Starobinsky model through the coupling to the trace anomaly \cite{Gorbunov:2012ns}:
\beq
\Omega \left< T^\mu_\mu \right> \; = \;\frac{\phi}{\sqrt{6}\,M_P}
\left< T^\mu_\mu \right>.\eeq
where $e^{2\Omega}$ is the conformal factor that transforms the theory from the Jordan to the Einstein frame, $\phi$ is the canonical inflaton and $T^\mu_\mu$ is the trace of the energy momentum tensor. This leads to a decay of the inflaton to gauge bosons \cite{Kamada:2019pmx}
\begin{align}
	\left.{T^\mu}_\mu\right\vert_\mathrm{anom} = \sum_{i,a} \frac{\beta_a}{4\alpha_a} F^{a}_{i\mu\nu} F_i^{a \mu\nu}\,,
 \label{anomT}
\end{align}
where $\beta_a$ is the beta function for the gauge group with generalized fine structure constant, $\alpha_a$ and field strength $F^a_{i\mu\nu}$ for $i=1-(N^2-1)$ for an SU(N) gauge group. This leads to a decay rate
\begin{align}
	\Gamma(\phi \to AA) = \sum_a \frac{N_{a} b_a^2 \alpha_a^2}{1536\pi^3}\frac{m^3}{M_P^2}\,,
 \label{anomD}
\end{align}
summed over groups with $N_a$ gauge bosons and leads to a reheating temperature of order $10^{8}$~GeV \cite{Ema:2024sit}. 

The connection between the $R+R^2$ theory of gravity and no-scale supergravity has been explored in some detail \cite{eno9,DLT, building,Antoniadis:2026uzn}, and supersymmetrization of the Starobinsky model
finds a natural formulation in the context of no-scale supergravity \cite{Cecotti}. As we discuss in more detail below, the no-scale K\"ahler potential with field space curvature $\mathcal{R} = 2/3$ and minimal kinetic terms for SM fields can be matched (in the Jordan frame) to fields conformally coupled to curvature. In this case, we expect, by analogy with the Starobinsky model, vanishing couplings of the inflaton to SM fields (up to terms such as the Higgs mass, which breaks the conformal invariance) leading to a very small decay rate for the inflaton and hence a very low reheating temperature \cite{Endo:2006xg,egno4}.

We discuss in this paper how this conclusion is avoided within well-motivated generalizations of the minimal no-scale realization of Starobinsky-like inflation. These include models in which the field space curvature $\mathcal{R}$ is modified via a change in the coefficient of the logarithm in the no-scale K{\"a}hler potential, as well as in models with non-minimal gauge kinetic terms and in the presence of anomaly-induced couplings. 

In what follows, we first review in Section~\ref{sec:no-scale} the construction of no-scale inflation models and their connections to $R+R^2$ gravity. Then, in Section~\ref{sec:infldecay} we derive the generalized couplings and decays of the inflaton to matter fields when the field space curvature is $\mathcal{R} = 2/3\alpha$, showing that only for the special case of $\alpha = 1$ do the tree-level couplings vanish. We then compute the reheating temperature as a function of $\alpha$ and explore the phenomenological consequences of this generalization, showing that the reheating temperature is greatly increased for generic values of $\alpha \ne 1$. We discuss in particular an explicit and motivated example, namely the string-derived model discussed in \cite{ANR1,ANR2,ANO}, showing that
it is indeed of the generalized form with $\alpha = 2/3$, and discuss the consequences for reheating in this model. In Section \ref{sec:T-gauge} we consider the couplings of the inflaton to gauge bosons (and gauginos) at the tree-level through a non-minimal gauge kinetic function and at one-loop order through the trace anomaly. We discuss the K{\"a}hler invariance of these couplings and the importance of this invariance in the context of inflationary avatars. We also present the predictions of the above scenarios in the $(n_s, r)$ plane.
For completeness, in Section \ref{sec:R3}, we also explore the role of a possible $R^3$ correction to the Einstein action and the impact it may have on reheating, which were recently considered in \cite{Antoniadis:2026uzn,Gialamas:2025ofz}.
Our conclusions are given in Section \ref{sec:summ}. 

\section{No-scale supergravity models of inflation}
\label{sec:no-scale}

Minimal no-scale supergravity models of inflation are described by two complex scalar fields $(T, C)$ in addition to gravity.
The corresponding no-scale supergravity theory is specified by the K\"ahler potential
\beq
K \; = \; - 3 \ln \left(T + {\bar T} - \frac{1}{3} C {\bar C} - \frac{1}{3} \sum_i |\phi_i|^2\right)  \, ,
\label{nsK}
\eeq
where SM particles are incorporated as untwisted fields $\phi_\alpha$. The
superpotential is given by
\beq
W = \sqrt{3} m C \left(T-\frac12 \right) + W_{\rm SM} (\phi_i) \, ,
\label{CSM}
\eeq
where $W_{\rm SM}(\phi_\alpha)$ is the SM superpotential. 
When the stabilizer field $C = 0$, the Lagrangian can be written as
\begin{eqnarray}
{\mathcal L} & = &   \frac{1}{12} (\partial_\mu K)^2 + e^{K/3} \left(|\partial_\mu C|^2 +|\partial_\mu \phi^i|^2\right) \nonumber \\
&& +\frac34 e^{2K/3} |\partial_\mu (T- T^*) + \frac13 (\phi_i^{*}\partial_{\mu}\phi^i-\phi^i\partial_{\mu}\phi^{*}_i)|^2  - e^{\frac23 K} {\hat V} \, ,
\label{LmanyJ}
\end{eqnarray}
and the scalar potential is given by 
\beq
V = e^{\frac23 K} {\hat V} =  e^{\frac23 K}   \left(\left|  W_{C} \right|^2 + \left|  W_{i} \right|^2  +\frac{1}{3} (T+T^*) |W_T|^2 +
\frac{1}{3} \left(W_T (\phi_i^* W^{*i} - 3 W^*) + {\rm h.c.}  \right) \right) \, .
\eeq
With $W_T = 0$,
the inflationary potential becomes
\beq
V = \frac{3 m^2 \left|(T-\frac12)\right|^2}{(T+{\bar T})^2} \, .
\eeq
Then, restricting to ${\rm Im}~T = 0$ and defining the canonical field $T \equiv \frac12 e^{\sqrt{\frac23} \frac{\phi}{M_P}}$, we obtain the Starobinsky potential in Eq.~(\ref{Staro}). 

It is sometimes convenient to represent the theory in a more symmetric basis.
Introducing the coordinates $y_0$ and $y$ given by
\begin{equation}
y_0
=
\frac{2C}{1+2T} \, ,
\qquad
y_i
=
\frac{2\phi_i}{1+2T} \, ,
\qquad
y
=
\sqrt{3}\,
\left( \frac{1-2T}{1+2T} \right) \, ,
\label{eq:Tphi-to-y}
\end{equation}
with the inverse relations
\begin{equation}
C
=
\frac{y_0}{1+y/\sqrt{3}}\, , \qquad
\phi_i
=
\frac{y_i}{1+y/\sqrt{3}}\, , \qquad
T
=
\frac12
\left( \frac{1-y/\sqrt{3}}{1+y/\sqrt{3}} \right) \, ,
\label{eq:y-to-Tphi}
\end{equation}
we can then rewrite the K\"ahler potential as 
\beq
K \; = \; - 3 \ln \left(1 - \frac13 |y|^2 - \frac13 |y_0|^2 - \frac13 \sum_i |y_i|^2 \right) \, .
\eeq
After the K\"ahler transformation that takes 
\beq
W(T, C, \phi_i) \; \to \; {\widetilde W}(y, y_0, y_i) \; = \; \left( 1 + {y}/{\sqrt{3}} \right)^3 W  \, ,
\label{Wtilde}
\eeq
the superpotential (\ref{CSM}) becomes
\beq
W = m y y_0 (1+y/\sqrt{3}) + W_{\rm SM}(y_i) \, .
\label{CSMs}
\eeq

The choice of the superpotential in Eq.~(\ref{CSM}) is not unique \cite{avatars}. Indeed, the first formulation of the Starobinsky model in the framework of no-scale supergravity~\cite{eno6} related the inflaton to $C$ rather than $T$, and postulated the Wess-Zumino superpotential 
\beq
W = m \left( \frac12 C^2 - \frac{1}{3\sqrt{3}} C^3 \right) \, .
\label{WWZ}
\eeq
This yields the Starobinsky potential when ${\rm{Im}}~C = 0$, $T = \frac12$, with the canonical inflaton field $\phi$ being defined by $C \equiv \sqrt{6} \tanh(\phi/\sqrt{6} M_P)$ \cite{eno6}. 

These models are part of a continuous set of possible models that all yield the Starobinsky potential and are related by the underlying SU(2,1)/SU(2)$\times$U(1) no-scale symmetry \cite{enov1,building}. 
At the quantum level, however, these models are not equivalent, as radiative corrections distinguish between them. 
For example, in the model described by
(\ref{WWZ}), supersymmetry is broken during inflation (at large $\phi$), and 1-loop corrections are significant \cite{egkko2}. For this reason, we focus hereafter on models more closely related to that defined by (\ref{CSM}), in which the inflationary potential does not induce large supersymmetry breaking and the 1-loop corrections are negligible.

The Lagrangian (\ref{LmanyJ}) derived from Eqs.~(\ref{nsK}) and (\ref{CSM}) can also be be derived from the $R+R^2$ model in the Jordan frame. 
Including a conformal coupling of the matter fields $\phi_i$ to curvature, we can start with the action
\begin{equation}
{\cal A} \; = \;  - \frac{1}{2} \int d^4x \sqrt{-g} \left[  R - \frac{1}{6 m^2} R^2 - 2 \left(\partial^\mu \phi^i \partial_\mu \phi^*_i  + \xi |\phi^i|^2 R - {\hat V} \right) \right] \, ,
\label{manyphi}
\end{equation}
we can introduce the Lagrange multiplier 
$ - R^2 \to 2 \Phi R + \Phi^2$ and perform a conformal transformation~\cite{WhittStelle,Kalara:1990ar}:
\begin{equation}
{\tilde g}_{\mu \nu}  \; = \; e^{2\Omega} g_{\mu \nu} \; = \;  \left(1 + \frac{1}{3 m^2} \Phi - 2 \xi  |\phi^i |^2\right) g_{\mu \nu} \, ,
\end{equation}
which leads to 
\begin{eqnarray}
\label{R2Einstein3}
{\cal A} \; & = & \;  - \frac{1}{2} \int d^4x \sqrt{-{\tilde g}} \left[ {\tilde R}  - 6 \partial^\mu \Omega \partial_\mu \Omega \right. \nonumber \\ 
&&
\left. \; - \;   2 e^{-2\Omega}   \partial^\mu \phi^i \partial_\mu \phi^*_i  + e^{-4\Omega} \left( \frac{1}{6 m^2}  \Phi^2 + 2 {\hat V} \right)  \right] \, ,
\end{eqnarray}
where ${\tilde R}$ is the curvature scalar in the Einstein frame.  With the identification 
\beq
\Omega = -K/6 \, ,
\eeq we recover the form of the Lagrangian in Eq.~(\ref{LmanyJ}) for $(1 + \frac{1}{3 m^2} \Phi) = T + {\bar T} $ and $\xi = 1/6$.~\footnote{Note that the additional terms in Eq.~(\ref{LmanyJ}) involving the imaginary parts of the scalar fields can be accounted for by including an auxiliary field, $b_\mu$ coupled to a current such that the Lagrangian includes $\frac13 b_\mu b^\mu -b_\mu J^\mu$ with $J^\mu = \frac13 (K_T \partial^\mu T + K_i \partial \phi^i - K^T \partial^\mu {\bar T} - K^i \partial^\mu \phi^*_i$) \cite{eno9}.  } For further details of this correspondence, see \cite{DLT,eno9,building}.

One of the great successes of the inflation paradigm is the possibility of generating nearly scale-independent density fluctuations. One can compare the theoretical predictions of the anisotropy spectrum with experiment by computing, for example, the tilt of the scalar power spectrum, $n_s$, its amplitude, $A_s$, and the ratio of the tensor-to-scalar amplitudes, $r$. These are conveniently calculated using the slow-roll parameters, $\epsilon$ and $\eta$.
For a given single-field scalar potential, these are given by
\begin{equation} 
\epsilon \; \equiv \; \frac{1}{2} M_{P}^2 \left( \frac{V'}{V} \right)^2 ; \; \;  \eta \; \equiv \; M_{P}^2 \left( \frac{V''}{V} \right)   \, ,
\label{epsilon}
\end{equation}
where, here and subsequently, the prime denotes a derivative with respect to the inflaton field $\phi$. The inflationary observables are then easily computed in terms of $\epsilon$ and $\eta$:
\begin{eqnarray}
A_s \;& = &\; \frac{V_*}{24 \pi^2 \epsilon_* M_{P}^4 } \simeq 2.1 \times 10^{-9} \, , \label{As} \\ \notag
 n_s \; & \simeq &\; 1 - 6 \epsilon_* + 2 \eta_* =  0.965 \pm 0.004 \; (68\%~{\rm CL}) \, ,
\label{ns} \\
 r \; &
\simeq & \; 16 \epsilon_* < 0.036 \; (95\%~{\rm CL}) \, . \label{r}
\label{observables}
\end{eqnarray}
In these expressions, the potential and slow-roll parameters are to be evaluated at the CMB pivot scale, $k_* = 0.05$~Mpc$^{-1}$. The experimental values for $A_s$ and $n_s$ are taken from {\it Planck} measurements~\cite{Planck} and the upper limit on $r$ from a combination of {\it Planck} and BICEP/Keck results~\cite{BICEP2021,Tristram:2021tvh}. We note that more recent results from the ACT collaboration~\cite{AtacamaCosmologyTelescope:2025blo} may indicate a higher value of $n_s$, which may call for deformations of the Starobinsky potential \cite{Antoniadis:2025pfa,Ellis:2025ieh,EGOV}, but that is not our focus here.~\footnote{Recent results from the SPT collaboration \cite{SPT-3G:2025bzu} are more consistent with those of Planck.}

The K\"ahler potential in Eq.~(\ref{nsK}) corresponds to a maximally-symmetric field space manifold with curvature $\mathcal{R} = 2/3$. As we indicated above, this can be mapped to the $R+R^2$ theory in the Jordan frame where SM fields are conformally coupled to (spacetime) curvature. 
In this case, however, as we have already seen, the inflaton decay channel to two Higgs bosons vanishes
when $\xi = 1/6$ up to a contribution from the coupling of the inflaton to the Higgs potential, which leads to a decay rate suppressed by $(m_H/M_P)^2$. 
In the no-scale framework, we expect (and will show below) that the couplings of the inflaton to matter either vanish or are exceedingly small \cite{Endo:2006xg,egno4,Ema:2024sit}. The curvature of the field space manifold can be easily altered while maintaining
a constant value by shifting the coefficient of the $\log$ in Eq.~(\ref{nsK}) from $-3$ to $-3\alpha$ in which case the curvature becomes $\mathcal{R} = 2/3\alpha$. As we show in the next Section, this has dramatic effects on the possible decay channels for the inflaton. 

A comprehensive study of the possible inflaton decay channels was performed in \cite{egno4}. 
Both $C$-type (referred to as an untwisted matter inflaton) and $T$-type inflation were considered. For $C$-type inflation, in the absence of a direct coupling of $C$ to SM fields in the superpotential, all tree-level decay channels vanish. However a coupling of the type $HLC$ in the superpotential (where $L$ denotes a lepton supermultiplet and $C$ is associated with a right-handed neutrino) can lead to a significant decay rate and sufficient reheating \cite{eno8}.~\footnote{There the association with the right-handed sneutrino was made to a related field in the symmetric basis where $T+{\bar T} - C {\bar C}/3 \to 1-y_0 {\bar y_0}/3 - y {\bar y}/3$. This basis will be discussed in more detail in Section \ref{sec:T-gauge}.  } Alternatively a modification of the gauge kinetic function that is dependent on $C$ can allow for decays to gauge bosons and gauginos \cite{Endo:2006xg,Kallosh:2011qk,egno4}. In contrast, when the inflaton is associated with the volume modulus $T$, tree-level couplings to the SM are non-zero. However, they typically lead to very small decay rates (and low reheating temperatures). For example, the decay to two Higgs bosons is present, but its coupling is suppressed by $(\mu_H/M_P)^2$, where $\mu_H$ is the supersymmetric bilinear Higgs mass term. For $\mu_H \sim 1$~TeV, the reheating temperature is only of order 0.1 eV.   Here we will restrict attention to $T$-type inflation. 
The study in \cite{egno4} was restricted to $\alpha = 1$. Viable models for inflation can be constructed in no-scale supergravity for $\alpha \ne 1$ \cite{EGOV} and impact of these models on reheating is discussed in the next Section.

\section{Inflaton decays and reheating}
\label{sec:infldecay}

At the tree level, one generally expects inflaton decays to all SM fields through supergravity couplings \cite{Endo:2006qk}. However, these couplings vanish for the special case of no-scale supergravity with a K\"ahler potential defined by Eq.~(\ref{nsK}) \cite{Endo:2006xg}. 
Here we extend the analysis in \cite{egno4}  to generalized no-scale models with
$\mathcal{R}=2/(3\alpha)$. For $\alpha = 1$,
the no-scale supergravity model can be seen as a supersymmetrization of the Starobinsky model (see the discussion in Section~\ref{sec:R3}). 
Although the direct correspondence with an $F(R)$ theory is generally lost for $\alpha\neq1$ (see the discussion in the Appendix), these models
provide the standard supergravity realization of $\alpha$-attractor
inflation. We show that the tree-level matter-coupling cancellation
present in the minimal $\alpha=1$ model is generally absent for
$\alpha\neq1$, although it may persist for special choices of the
matter-field modular weights.

We take the visible-sector K\"ahler potential in Planck units to be
\begin{equation}
K
\;=\;
-3\alpha \,
\ln\!\left[
T+\bar T
\;-\;
\frac{1}{3}\sum_i |\phi_i|^{2}
\right]
\;+\;
\sum_a
\frac{|\varphi_a|^{2}}{(T+\bar T)^{n_a}}\,,
\label{eq:K-alpha-decays}
\end{equation}
where the fields $\phi_i$ are untwisted matter fields appearing inside the no-scale logarithm, and the $\varphi_a$ are twisted fields with modular weights $n_a$. The stabilizing field $C$ can be associated with one of the $\phi_i$. e.g., $\phi_0$.  The dilaton $S$ and the remaining moduli are assumed to be stabilized at a higher scale and decoupled from the inflationary dynamics. 

Starobinsky-like inflation can be obtained for $\alpha \ne 1$ assuming a superpotential of the form \cite{EGOV}
\beq
W= A \sqrt{\lambda} M_P^2 C \left(\frac{2T}{M_P}\right) f\left(\frac{T}{M_P}\right) \, ,
\label{simpT}
\eeq
with the specific choice 
\beq
f(T/M_P) = \left(1-\frac{M_P}{2T} \right) \left(\frac{T}{M_P} \right)^{(3 \alpha -3)/2} \, ,
\label{femodel}
\eeq
and $A^2 = 3 \cdot 2^{(3\alpha-5)} \alpha$.
Using Eq.~(\ref{simpT}), the general form of the potential is 
\beq
V = 4 A^2 \lambda M_P^4 \frac{|T/M_P|^2 |f(T/M_P)|^2}{\alpha (T/M_P+{\bar T}/M_P)^{3\alpha-1}} \, ,
\eeq
when one fixes $C = 0$. For $f(T)$ given in Eq.~(\ref{femodel}), we recover the Starobinsky-like potential~\cite{e-m}:
\begin{align}
V & \; = \; \frac{3}{4} \lambda  M_{P}^{4}\left(1-e^{-\sqrt{\frac{2}{3 \alpha}} \frac{\phi}{M_{P}}}\right)^{2} \, ,
\label{eq:emodel}
\end{align}
where
the canonical inflaton is given by 
\begin{equation}
\label{eq:tcanonicalfield}
T = \bar{T} = \frac{1}{2}e^{\sqrt{\frac{2}{3\alpha}} \frac{\phi}{M_P}}\,.
\end{equation}

We now discuss the perturbative decays of the inflaton when it is identified with the real part of the volume modulus $T$. 
The original analysis of the decay channels at the $\alpha=1$ no-scale point was given in Ref.~\cite{egno4}. Here we generalize the analysis to arbitrary $\alpha$. In the next Section we will apply these results to the ANR model, for which $\alpha_{\rm ANR}=2/3$.

As noted above, in the $\alpha=1$ no-scale limit, the modulus does not have unsuppressed tree-level decays into untwisted matter fields~\cite{egno4}. The same cancellation occurs for twisted fields whose modular weights reproduce the conformal value $n_I+n_J+n_K=3$ (see the discussion and expressions below). This is
the supergravity counterpart of the vanishing of the scalaron decay amplitude
into conformally coupled scalars in the Jordan-frame description of the $R+R^2$ model~\cite{eno9,building}. The cancellation may be lifted by departing from $\alpha=1$, by introducing matter with modular weights $n_a\neq 1$, by adding a nontrivial gauge kinetic function, or by the 
super-Weyl anomalies that generate radiative couplings of $T$ to the gauge sector. We address each of these in turn.

At $\phi_i=\varphi_a=0$, the modulus kinetic term is
\begin{equation}
{\cal L}_{\rm kin}
\;=\;
-\,3\alpha M_P^2\,
\frac{\partial_\mu T\,\partial^{\mu}\bar T}{(T+\bar T)^{2}}\,.
\end{equation}
Along the real trajectory we introduce the canonically-normalized real
modulus fluctuation $\delta T$ via
\begin{equation}
T+\bar T
\;=\;
\exp\!\left[
\sqrt{\frac{2}{3\alpha}}\,\delta T
\right]\,,
\qquad
\langle T+\bar T\rangle=1\,,
\label{eq:T-canonical-decay}
\end{equation}
so that, linearizing around the vacuum,
\begin{equation}
T+\bar T
\;=\;
1+\sqrt{\frac{2}{3\alpha}}\,\delta T
+\frac{1}{3\alpha}\delta T^{\,2}
+\cdots .
\label{eq:TpTbar-exp}
\end{equation}
Throughout this Section, $\delta T$ denotes the canonically-normalized real modulus fluctuation in reduced Planck units. The physical inflaton mass at the vacuum is denoted $m_{\delta T}\equiv m= \sqrt{\lambda/\alpha} M_P$.

Expanding (\ref{eq:K-alpha-decays}) at the vacuum shows that untwisted matter fields acquire an effective modular weight $n=1$ after canonical rescaling. Indeed, at quadratic order in matter,
\begin{equation}
K
\;\supset\;
\frac{\alpha\,\phi_i\bar\phi_i}{T+\bar T}
\;+\;\cdots\,,
\end{equation}
so after the constant rescaling $\phi_i^{\,c}=\sqrt{\alpha}\,\phi_i$, the matter K\"ahler metric reduces to
\begin{equation}
K_{i{\bar i}} \equiv Z_i(T,\bar T)
\;=\;
(T+\bar T)^{-1}\, ,
\end{equation}
where $K_{i{\bar i}} = \partial^2 K/\partial \phi_i \partial {\bar \phi}_i$.
Thus, as far as the $T$-dependence of decay amplitudes is concerned,
untwisted matter is accounted for by setting
\begin{equation}
n_I\;=\;1\,,
\end{equation}
i.e., identical to a twisted field with unit modular weight. This identification
holds independently of $\alpha$ and is the natural extension of the
$\alpha=1$ rule of Ref.~\cite{egno4}.

The visible sector superpotential is assumed to contain ordinary matter couplings,
\begin{equation}
W_{\rm vis}
\;=\;
\frac{1}{2}\,\mu^{IJ}\Phi_{I}\Phi_{J}
\,+\,
\frac{1}{6}\,Y^{IJK}\Phi_{I}\Phi_{J}\Phi_{K}
\,+\,\cdots\,,
\label{eq:Wvis-decays}
\end{equation}
where $\Phi_I$ denotes either an untwisted or twisted visible multiplet, and
the modular weight $n_I$ of each $\Phi_I$ is fixed as discussed above. We
assume that $W_{\rm vis}$ has no explicit dependence on $T$, which is the
sequestered case relevant for the no-scale cancellation. Explicit
$T$-dependence in $W_{\rm vis}$ would generate additional model-dependent
decay channels along the lines of Eq.~(5.50) of Ref.~\cite{egno4}, and is not included in the minimal rates below.

\subsection{Effective interaction Lagrangians}
\label{sec:T-Leff}
After canonically normalizing the matter fields at the vacuum and expanding in $\delta T$, we obtain the linear in $\delta T$ effective interaction Lagrangian to all orders in the matter fields. The fermion sector reads 
\begin{align}
{\cal L}_{F,{\rm eff}}
&=
-\,\frac{\delta T}{2\sqrt{3\alpha}}
\bigl(n_I+n_J-3\alpha\bigr)\,W^{IJ}\,
\bar\chi_{I,L}\chi_{J,R}
\nonumber\\
&\quad
-\,\frac{\delta T}{2\sqrt{3\alpha}}
\bigl(n_I+n_J+n_K-3\alpha\bigr)\,W^{IJK}\,
\bar\chi_{I,L}\chi_{J,R}\Phi_K
\;+\;{\rm h.c.}+\cdots ,
\label{eq:LF-eff-alpha}
\end{align}
and the bosonic sector is
\begin{align}
\mathcal{L}_{B,\mathrm{eff}}
=&
-\frac{\delta T}{\sqrt{3\alpha}}\,
\left(n_I+n_L-3\alpha\right)
W^{IL}\bar W_{LJ}\Phi_I\bar\Phi^J
\nonumber\\[0.5em]
&-\frac{\delta T}{2\sqrt{3\alpha}}\,
\left(n_I+n_L-3\alpha\right)
W^{IL}\bar W_{LJK}\Phi_I\bar\Phi^J\bar\Phi^K
\nonumber\\[0.5em]
&-\frac{\delta T}{2\sqrt{3\alpha}}\,
\left(n_I+n_J+n_L-3\alpha\right)
W^{IJL}\bar W_{LK}\Phi_I\Phi_J\bar\Phi^K
\nonumber\\[0.5em]
&-\frac{\delta T}{\sqrt{3\alpha}}\,
\left(n_J+n_L-3\alpha\right)
W^{JL}\bar W_{LK}\Phi_I\Phi_J\bar\Phi^I\bar\Phi^K
\nonumber\\[0.5em]
&-\frac{\delta T}{6\sqrt{3\alpha}}\,
\left(n_I+n_L-3\alpha\right)
W^{IL}\bar W_{LJKM}\Phi_I\bar\Phi^J\bar\Phi^K\bar\Phi^M
\nonumber\\[0.5em]
&-\frac{\delta T}{4\sqrt{3\alpha}}\,
\left(n_I+n_J+n_L-3\alpha\right)
W^{IJL}\bar W_{LKM}\Phi_I\Phi_J\bar\Phi^K\bar\Phi^M
\nonumber\\[0.5em]
&-\frac{\delta T}{6\sqrt{3\alpha}}\,
\left(n_I+n_J+n_K+n_L-3\alpha\right)
W^{IJKL}\bar W_{LM}\Phi_I\Phi_J\Phi_K\bar\Phi^M 
\nonumber\\[0.5em]
&-\frac{\delta T}{12\sqrt{3\alpha}}\,
\left(n_I+n_J-3\alpha\right)
\left[
9\alpha+
\left(n_I+n_J-1\right)
\left(n_K+n_M-3\alpha\right)
\right]
W^{IJ}\bar W_{KM}\Phi_I\Phi_J\bar\Phi^K\bar\Phi^M \label{eq:LB-eff-alpha} \\
& +\cdots \nonumber.
\end{align}
The displayed terms generate the two-, three-, and four-body scalar decay channels, and generalize the eight operators of Eq.~(5.40) of Ref.~\cite{egno4} from $\alpha=1$ to arbitrary $\alpha$, with the universal coefficient combinations
\begin{equation}
n_I+n_J-3\alpha\,,
\qquad
n_I+n_J+n_K-3\alpha\,,
\qquad
n_I+n_J+n_K+n_L-3\alpha\,,
\label{eq:no-scale-combinations}
\end{equation}
which encode the residual no-scale cancellation.

For $\alpha=1$ and matter fields with modular weight $n_I=1$, every combination in (\ref{eq:no-scale-combinations}) reduces to the form $n_I+n_J+\cdots+n_L-3$ in \cite{egno4}, and in particular vanishes for a trilinear Yukawa interaction among three untwisted fields:
\begin{equation}
\bigl(n_I+n_J+n_K-3\bigr)\big|_{n=1}=0\,.
\end{equation}
This recovers the conformal cancellation noted in Eq.~(5.40) of Ref.~\cite{egno4}. Twisted fields with modular weights arranged such that the relevant sum equals $3\alpha$ are similarly suppressed.

\subsection{Decays to matter scalars}
\label{sec:T-scalars}
We next examine in more detail the possible decay modes for the inflaton, beginning with decays to matter scalars. 
The two-body scalar decay $\delta T \to \Phi_I\bar\Phi^J$ proceeds via the first term of (\ref{eq:LB-eff-alpha}). Assuming negligible final-state masses relative to $m$, the partial width is
\begin{equation}
\Gamma\bigl(\delta T \to \Phi_I\bar\Phi^J\bigr)
\;=\;
\frac{\bigl(n_I+n_L-3\alpha\bigr)^{2}}{\alpha}
\frac{\bigl|W^{IL}\bar W_{LJ}\bigr|^{2}}{48\pi\,m\,M_{P}^{2}}\,.
\label{eq:Gamma-2body-scalar}
\end{equation}
As discussed in \cite{egno4}, this rate is proportional to the fourth power of the
bilinear coupling: for a Higgsino mass term $W \supset \mu_H H_uH_d$ it
yields a contribution suppressed by $|\mu_H|^{4}/(mM_{P}^{2})$ and is
phenomenologically negligible for $|\mu_H|\ll m$. Nevertheless, this channel can be important in models of high-scale supersymmetry \cite{Dudas:2017kfz,Kaneta:2019yjn}.

The three-body decays into scalars come in two topologies:
\begin{align}
\Gamma\bigl(\delta T\to \Phi_I\bar\Phi^{J}\bar\Phi^{K}\bigr)
&=
\frac{(n_I+n_L-3\alpha)^{2}}{\alpha}\,
\frac{\bigl|W^{IL}\bar W_{LJK}\bigr|^{2}\,m}{12(8\pi)^{3}\,M_{P}^{2}}\,,
\label{eq:Gamma-3body-scalar-A}\\[4pt]
\Gamma\bigl(\delta T\to \Phi_I\Phi_J\bar\Phi^{K}\bigr)
&=
\frac{(n_I+n_J+n_L-3\alpha)^{2}}{\alpha}\,
\frac{\bigl|W^{IJL}\bar W_{LK}\bigr|^{2}\,m}{12(8\pi)^{3}\,M_{P}^{2}}\,,
\label{eq:Gamma-3body-scalar-B}
\end{align}
each of which involves the insertion of one bilinear superpotential term and one trilinear term.
These channels are suppressed by the square of the
bilinear coupling and lead to MeV reheating temperatures for $\mu_H$ of order 1 TeV. These are subleading relative to the corresponding four-scalar mode below,
which is unsuppressed by visible-sector masses.

The four-scalar decay rate, dominated by the trilinear-trilinear coupling,
is
\begin{equation}
\Gamma\bigl(\delta T \to \Phi_I\Phi_J\bar\Phi^{K}\bar\Phi^{M}\bigr)
\;=\;
\frac{(n_I+n_J+n_L-3\alpha)^{2}}{\alpha}\,
\frac{\bigl|W^{IJL}\bar W_{LKM}\bigr|^{2}\,m^{3}}{72(8\pi)^{5}\,M_{P}^{2}}\,,
\label{eq:Gamma-4body-scalar}
\end{equation}
which generalizes Eq.~(5.46) of \cite{egno4} for general $\alpha$. This channel can provide reheating temperatures as large as $\mathcal{O}(10^7)$~GeV so long as $n_I + n_J + n_K - 3\alpha \ne 0$. However, for $\alpha = 1$ and untwisted matter fields (or twisted matter fields with $n_i = 1$), this rate also vanishes. Because this is a four-body channel, this width is suppressed relative to the three-body rate with fermions in the final state.

\subsection{Decays to matter fermions}
\label{sec:T-fermions}

The two-body fermion decay $\delta T \to \bar\chi_I\chi_J$ is mediated by
the first term of (\ref{eq:LF-eff-alpha}) and is the dominant bilinear
channel. Once again neglecting the final state masses, the decay width for this channel is
\begin{equation}
\Gamma\bigl(\delta T \to \bar\chi_I\chi_J\bigr)
\;=\;
\frac{(n_I+n_J-3\alpha)^{2}}{\alpha}\,
\frac{|W^{IJ}|^{2}\,m}{192\pi\,M_{P}^{2}}\,.
\label{eq:Gamma-2body-fermion}
\end{equation}
For a Higgsino mass term $W\supset \mu_H H_uH_d$, taking the Higgses to be
untwisted (so $n_{H_u}=n_{H_d}=1$), this becomes
\begin{equation}
\Gamma\bigl(\delta T \to \widetilde H_u^{0}\widetilde H_d^{0}\bigr)
\;=\;
\frac{(2-3\alpha)^{2}}{\alpha}\,
\frac{|\mu_H|^{2}\,m}{192\pi\,M_{P}^{2}}\,.
\label{eq:Gamma-mu-alpha}
\end{equation}
If $\alpha=1$ the coefficient $(2-3\alpha)^{2}/\alpha=1$ does not
vanish, but the rate is suppressed by $|\mu_H|^{2}/m^{2}$ relative to the
unsuppressed gravitational scale $m^{3}/M_P^{2}$. It therefore provides only
a small reheating floor (of order 10 MeV) and is subdominant to the trilinear channels in any
realistic spectrum.

The dominant fermionic decay is the three-body trilinear decay
$\delta T \to \bar\chi_I\chi_J\Phi_K$, with partial width
\begin{equation}
\Gamma\bigl(\delta T \to \bar\chi_I\chi_J\Phi_K\bigr)
\;=\;
\frac{(n_I+n_J+n_K-3\alpha)^{2}}{\alpha}\,
\frac{|W^{IJK}|^{2}\,m^{3}}{36(8\pi)^{3}\,M_{P}^{2}}\,.
\label{eq:Gamma-3body-fermion}
\end{equation}
The corresponding scalar three-body widths (\ref{eq:Gamma-3body-scalar-B})
and (\ref{eq:Gamma-4body-scalar}) combine with this fermion mode into the
total three-body Yukawa channel.
In the case of the top Yukawa coupling, and taking the three fields to be untwisted with
$n_{Q_3}=n_{H_u}=n_{\bar u_3}=1$, the relevant coefficient in brackets is
\begin{equation}
n_{Q_3}+n_{H_u}+n_{\bar u_3}-3\alpha
\;=\;
3(1-\alpha)\,,
\label{eq:top-coeff-alpha}
\end{equation}
giving the combined three-body partial width
\begin{equation}
\Gamma(\delta T \rightarrow H_u^0t_L\bar{t}_R, \,  \tilde{t}_L\tilde{H}_u^0\bar{t}_R, \, \bar{\tilde{t}}_R t_L\tilde{H}_u^0)
\;=\;
\frac{3(1-\alpha)^{2}}{\alpha}\,
\frac{|y_t|^{2}\,m^{3}}{4(8\pi)^{3}\,M_{P}^{2}}\,.
\label{eq:Gamma-top-alpha}
\end{equation}
The rate (\ref{eq:Gamma-top-alpha}) vanishes at $\alpha=1$, in agreement with
\cite{egno4}, but is otherwise determined entirely by the K\"ahler geometry once the
matter assignments are fixed. Equation (\ref{eq:Gamma-top-alpha}) is the
central result of this Section for $\alpha\neq 1$. This channel can produce reheating temperatures of order $10^8$~GeV.

\subsection{Reheating}
\label{sec:reheating}
The predictions for the CMB observables depend not only on the inflationary potential, but also on the post-inflationary expansion history. In particular, the value of the inflaton field at horizon exit is fixed by the number of e-folds $N_*$ between the exit from the pivot mode and the end of inflation, and $N_*$ depends logarithmically on the duration of reheating. We follow the treatment of~\cite{EGOV}, replacing the phenomenological inflaton decay rate used there by the decay rate derived from the no-scale matter couplings derived in the previous subsections.

For the $\alpha$-Starobinsky potential (\ref{eq:emodel}), the inflaton mass at the minimum is $m = \sqrt{\frac{\lambda}{\alpha}}M_P.$ The normalization $\lambda$ is fixed for every value of $\alpha$ by the measured scalar amplitude, as in Eq.~(\ref{As}). A good approximation for $\lambda$ was given in \cite{Ellis:2021kad}
\beq
\lambda \simeq \frac{24 \alpha \pi^2 A_s}{N_*^2} \, .
\label{lapprox}
\eeq
Thus, to leading order in the attractor expansion, the explicit $\alpha$-dependence cancels from the inflaton mass. The principal $\alpha$-dependence of reheating therefore arises from the decay amplitude rather than from $m$, although the exact numerical result retains a mild additional dependence through the CMB normalization and the finite-$N_*$ dynamics.

Using the results from the previous subsection, the dominant decay channel is into the scalar--fermion final states associated with the top Yukawa coupling, given by Eq.~(\ref{eq:Gamma-top-alpha}).
Compared with the phenomenological parametrization
\begin{equation}
    \Gamma_\phi \; = \; 
\frac{y_{\rm eff}^2}{8\pi}\,m 
\label{eq:y-effective-definition}
\end{equation}
used in \cite{EGOV}, Eq.~\eqref{eq:Gamma-top-alpha} corresponds to
\begin{equation}
y_{\rm eff}^{\rm top}
=
\frac{\sqrt{3}}{16\pi}
\frac{|1-\alpha|}{\sqrt{\alpha}}\,
|y_t|\frac{m}{M_P}
\label{eq:y-effective-top}
\end{equation}
when $\alpha \ne 1$.
At $\alpha = 1$, however, Eq.~\eqref{eq:Gamma-top-alpha} vanishes exactly, and the reheating history is instead determined by the anomaly--induced decays to gauge bosons and gauginos (see Eq.~(\ref{anomD}) and its no-scale analogue in Section \ref{sec:T-gauge}). 

Perturbative reheating can be described by
the coupled equations \cite{GKMO2}
\begin{align}
\dot\rho_\phi
+
3H\bigl(1+w_\phi\bigr)\rho_\phi
&=
- \bigl(1+w_\phi\bigr) \Gamma_\phi\rho_\phi \, ,
\label{eq:rho-phi-reheating}
\\
\dot\rho_R+4H\rho_R
&=
\bigl(1+w_\phi\bigr) \Gamma_\phi\rho_\phi \, ,
\label{eq:rho-R-reheating}
\\
3M_P^2H^2
&=
\rho_\phi+\rho_R \, .
\label{eq:Friedmann-reheating}
\end{align}
Near its minimum the potential in
Eq.~\eqref{eq:emodel} is quadratic. The coherently
oscillating inflaton condensate therefore has $\langle w_\phi\rangle\simeq0$, and behaves as nonrelativistic matter until its decay products dominate. Initially, $\rho_\phi\propto a^{-3}$, whereas the continuously
produced radiation approaches the familiar scaling
$\rho_R\propto a^{-3/2}$, or $T\propto a^{-3/8}$, during the inflaton-dominated phase \cite{Scherrer:1984fd,Giudice:2000ex,GKMO1,GKMO2}.

We define the completion of reheating when the energy density in radiation is equal to that in the inflaton condensate,
for which the reheating temperature is given by Eq.~(\ref{TRH}). 
Equivalently,
\begin{equation}
\trh
=
\left(
\frac{72}{5\pi^2g_{\rm RH}}
\right)^{1/4}
\sqrt{\Gamma_\phi M_P} \, .
\label{eq:Treh-Gamma}
\end{equation}
When the top channel dominates, and we sum over the three final states and include color factors, this gives 
\begin{align}
\trh^{\rm top}
={}&
\left(
\frac{72}{5\pi^2g_{\rm RH}}
\right)^{1/4}
\left[
\frac{
\left(
n_{Q_3}
+n_{H_u}
+n_{\bar u_3}
-3\alpha
\right)^2
}{4\alpha}
\right]^{1/2}\times
\frac{|y_t|\,m^{3/2}}
{(8\pi)^{3/2}M_P^{1/2}} \, .
\label{eq:Treh-top-general}
\end{align}
For $n_i = 1$, the parametric dependence for $\trh$ becomes
\begin{equation}
\trh^{\rm top}
\propto
\frac{|1-\alpha|}{\sqrt{\alpha}}\,
|y_t|\,\frac{m^{3/2}}{M_P^{1/2}} \, .
\label{eq:Treh-alpha-scaling}
\end{equation}
Since $m$ is approximately independent of $\alpha$ after imposing the CMB normalization, this expression makes the dominant geometrical dependence explicit.

Assuming no entropy production after reheating, the number of e-folds at the pivot scale is~\cite{LiddleLeach,Martin:2010kz,Ellis:2021kad,EGOV}
\begin{align}
N_*
={}&
\ln\left[
\frac{1}{\sqrt{3}}
\left(\frac{\pi^2}{30}\right)^{1/4}
\left(\frac{43}{11}\right)^{1/3}
\frac{T_0}{H_0}
\right]
-
\ln\left(\frac{k_*}{a_0H_0}\right)
-
\frac{1}{12}\ln g_{\rm RH}
\nonumber\\
&+
\frac14
\ln\left(
\frac{V_*^2}{M_P^4\rho_{\rm end}}
\right)
+
\frac{1-3w_{\rm int}}
{12(1+w_{\rm int})}
\ln\left(
\frac{\rho_{\rm rad}}{\rho_{\rm end}}
\right) \, ,
\label{eq:Nstar-reheating-matching}
\end{align}
where $\rho_{\rm end} = \frac32 V(\phi_{\rm end})$ is the energy density at the end of inflation,
$\rho_{\rm rad} = \rho_{\rm R}(\trh)$ is the energy density at the onset of radiation
domination, and
\begin{equation}
w_{\rm int}
\equiv
\frac{1}{N_{\rm rad}-N_{\rm end}}
\int_{N_{\rm end}}^{N_{\rm rad}}
w(N)\,dN
\label{eq:wint-definition}
\end{equation}
is the equation-of-state parameter averaged over reheating. For the quadratic minimum considered here, $w_{\rm int}$ is close to zero, but we determine it together with $\rho_{\rm rad}$ by solving Eqs.~\eqref{eq:rho-phi-reheating}--\eqref{eq:Friedmann-reheating}.

For each value of $\alpha$, we begin with an initial estimate of $N_*$ and determine the corresponding horizon-exit field value
$\phi_*$ from
\begin{equation}
\label{eq:efolds}
N_{*} \; = \; \int_{t_*}^{t_{\rm end}} H \, dt \; \simeq \; \frac{1}{M_{P}^{2}} \int_{\phi_{\mathrm{end}}}^{\phi_{*}} \frac{V(\phi)}{V'(\phi)} d \phi 
\; \simeq \; \int_{\phi_{\mathrm{end}}}^{\phi_{*}} \frac{1}{\sqrt{2 \epsilon}} \frac{d \phi}{M_{P}}\,,
\end{equation}
We then fix $\lambda$ from the observed scalar
amplitude $A_s$ using Eq.~\eqref{As} (or the approximation given in Eq.~(\ref{lapprox}), compute the inflaton
mass $m$ and the top-Yukawa decay width $\Gamma_{\rm top}(\alpha)$ from Eq.~\eqref{eq:Gamma-top-alpha}, and evolve the coupled inflaton and radiation densities through reheating. From this evolution, we determine $w_{\rm int}$ and $\rho_{\rm rad}$, update $N_*$ using Eq.~\eqref{eq:Nstar-reheating-matching}, and repeat the procedure until $N_*$ converges.
Once $N_*$ and $\phi_*$ are known, the inflationary observables, $n_s$ and $r$ can be computed using Eq.~(\ref{ns}).

\begin{figure}[ht!]
    \centering
    \includegraphics[width=0.8\linewidth]{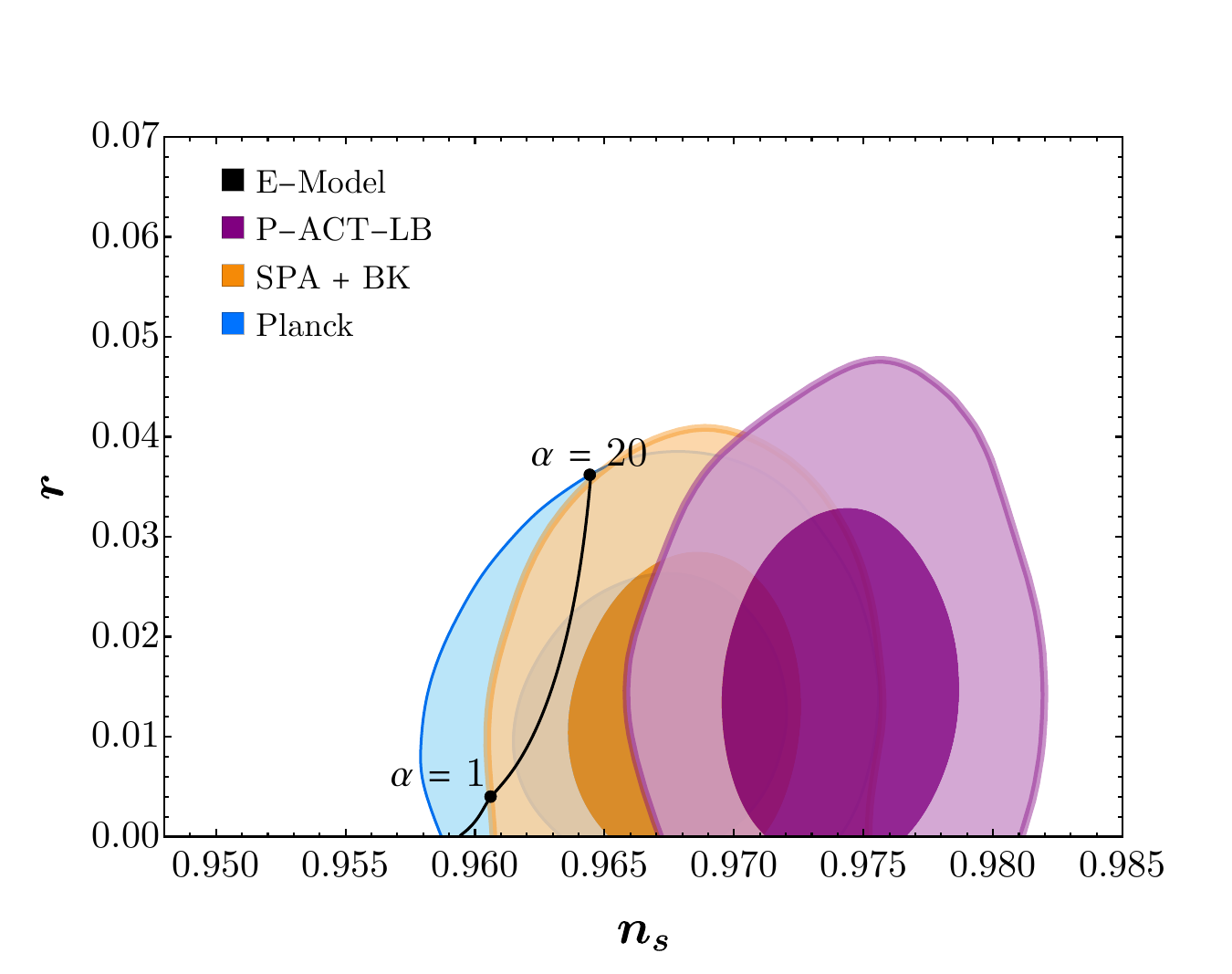}
    \caption{Predictions for the CMB observables $(n_s,r)$ in the generalized no-scale supergravity model of Eq.~\eqref{eq:K-alpha-decays}, shown as a function of $\alpha$ (black line). The shadings indicate the regions of parameter space favored  at the 68 and 95\% C.L. by {\it Planck} \cite{Planck}, the combination of {\it Planck}, ACT lensing, and DESI DR6 BAO data (P-ACT-LB) \cite{AtacamaCosmologyTelescope:2025blo}, and the combination of SPT, {\it Planck}, ACT and BICEP/Keck data \cite{SPT-3G:2025bzu,Balkenhol:2025wms} (labeled SPA + BK). The points corresponding to $\alpha=1$ and $\alpha=20$ are indicated explicitly.}
    \label{fig:nsrplot1}
\end{figure}

This produces the curve in the $(n_s,r)$ plane shown in
Fig.~\ref{fig:nsrplot1}. Each point on the black line corresponds to a different value of $\alpha$, with the normalization and reheating history calculated as described above. The point $\alpha=1$ requires special treatment because the top-Yukawa decay width vanishes there. Its location is evaluated using the anomaly--induced decay channel given by Eq.~(\ref{anomD}) and discussed in Section~\ref{sec:T-gauge}. 
For $\alpha\neq1$, the top-Yukawa channel provides the direct link between the no-scale K\"ahler geometry, the reheating temperature, and the location of the inflationary prediction in the observational $(n_s,r)$ plane.

We emphasize one key difference between this result and the recent analysis in \cite{EGOV}. In the latter, $\alpha$-attractor models with a potential given by Eq.~(\ref{eq:emodel}) were considered and the inflationary observables were computed as a function of $\alpha$ and independently $\trh$. This resulted in a series of (nearly parallel) trajectories with varying $\alpha$ for each assumed reheating temperature. However, in the context of no-scale supergravity, we see that the reheating 
is determined by $\alpha$ from Eq.~(\ref{eq:Treh-alpha-scaling}) for $\alpha \ne 1$, or by the decay of the inflaton to gauge bosons through the coupling of the trace anomaly for $\alpha = 1$. This results in the single trajectory shown in Fig.~\ref{fig:nsrplot1}.

For the range
\begin{equation}
0.01 \leq \alpha \leq 20,
\end{equation}
the corresponding reheating temperatures span approximately
\begin{equation}
2.3\times10^{8}\ {\rm GeV}
\lesssim
\trh
\lesssim
2.5\times10^{11}\ {\rm GeV},
\end{equation}
with the minimum occurring near $\alpha=1$, where reheating is
controlled by the trace-anomaly channel, and the maximum occurring when $\alpha = 0.01$. The points $\alpha=1$ and $\alpha=20$ are indicated explicitly in Fig.~\ref{fig:nsrplot1}.

The trajectory in Fig.~\ref{fig:nsrplot1} is compared with recent experimental results.
We note that the no-scale model prediction is consistent with the {\it Planck} data~\cite{Planck} (shown by the blue shaded region) for $\alpha \le 20$ at the 95\% C.L.. For a limited range of $\alpha > 1$, the model is also consistent with {\it Planck} at the 68\% C.L. shown by the dark blue shaded region.  The model is consistent with the SPT~+~BK combination~\cite{SPT-3G:2025bzu,Balkenhol:2025wms} (labelled as SPA+BK since it includes Planck and ACT data, and shaded orange) for $1 \le \alpha \le 20$. However, the no-scale prediction lies outside the 95\% C.L. region of the P-ACT-LB combination~\cite{AtacamaCosmologyTelescope:2025blo} (shaded purple) for all values of $\alpha$.

\section{The ANR model}
\label{sec:anr}
We now apply the general results of Section~\ref{sec:reheating} to the string-derived model of Refs.~\cite{ANR1,ANR2,ANO}. This model provides an explicit realization of no-scale inflation with $\alpha_{\rm ANR}=2/3$. As we will see, the tree-level no-scale cancellations found at $\alpha=1$ are present in this model as well. However, the model contains a direct coupling of the inflaton to matter allowing for successful reheating. 

In the symmetric field basis, the relevant string K\"ahler potential is
\begin{equation}
K_{\rm string}
=
-\ln(S+\bar S)
-2\ln\left(1-|y|^2\right)
-2\ln\left(1-\frac{|z|^2}{2}\right) \, ,
\label{eq:ANR-K-yz}
\end{equation}
and the superpotential is 
\begin{equation}
W_{\rm inf}
=
mzy(1- {\hat \lambda} y) \, 
\label{eq:ANR-W-yz}
\end{equation}
where ${\widehat \lambda}$ is a coupling of $\mathcal{O}(1)$.  
Here $S$ is the string dilaton, $y$ contains the inflaton, and $z$ contains the goldstino during inflation. This is related to the superpotential in Eq.~(\ref{CSMs}) with the identification of $z$ with $C$ and setting $\widehat \lambda = 1$. Departures from this value of $\widehat \lambda$ will produce deformations of the Starobinsky-like potential which may lead to interesting consequences for the inflationary observables, $n_s$ and $r$ \cite{Antoniadis:2025pfa}.  We assume that the dilaton is stabilized and absorb its vacuum expectation value into the definition of the mass parameter $m$. This assumption is sufficient for the discussion of reheating, whose relevant modulus dependence resides in the $y$, or equivalently $T$, sector.
We note that the stabilization of $S$ naturally leads to the stabilization of $C$, as is necessary for inflation \cite{ANO}. 

The supergravity scalar potential
takes the form (assuming $\hat \lambda = 1$)
\begin{align}
V
={}&
\frac{m^2}{
\left(1-|y|^2\right)^2
\left(2-|z|^2\right)^2}
\Bigl[
4|y|^2|1-y|^2
\nonumber\\
&\quad
+2|z|^2
\left\{
1-2(y+\bar y)+2|y|^2
+2(y+\bar y)|y|^2-3|y|^4
\right\}
\nonumber\\
&\quad
+|y|^2|1-y|^2|z|^4
\Bigr] \, .
\label{eq:ANR-full-potential}
\end{align}
Along the inflationary trajectory,
\begin{equation}
{\rm Im}\,y=0 \, ,
\qquad
z=0 \, ,
\end{equation}
and this expression reduces to 
\begin{equation}
V(y_R)
=
m^2\frac{y_R^2}{(1+y_R)^2} \, ,
\qquad
y_R\equiv {\rm Re}\,y \, .
\label{eq:ANR-potential-y}
\end{equation}
Consequently,
\begin{equation}
V(\phi)
=
\frac{m^2M_P^2}{4}
\left(
1-e^{-\phi/M_P}
\right)^2 \, ,
\label{eq:ANR-Starobinsky}
\end{equation}
where $\phi$ is again the canonically normalized inflaton.
This is equivalent to the $\alpha$-Starobinsky potential in Eq.~(\ref{eq:emodel})
with $\alpha=2/3$.

In the $(T,C)$ basis,
after the associated K\"ahler transformation, the inflationary sector may be
written as
\begin{equation}
K_{\rm string}
=
-\ln(S+\bar S)
-2\ln(T+\bar T)
-2\ln\left(1-\frac{|z|^2}{2}\right) \, ,
\label{eq:ANR-K-Tz}
\end{equation}
with
\begin{equation}
W_{\rm inf}
=
mz\left(T-\frac{1}{2}\right) \, .
\label{eq:ANR-W-Tz}
\end{equation}
The coefficient of the modulus logarithm is therefore $-2=-3\alpha_{\rm ANR}$, which makes the identification
$\alpha_{\rm ANR}=2/3$ manifest.

At $z=0$, the modulus kinetic term is
\begin{equation}
{\cal L}_{\rm kin}
=
-2M_P^2
\frac{\partial_\mu T\partial^\mu\bar T}
{(T+\bar T)^2} \, .
\label{eq:ANR-T-kinetic}
\end{equation}
Along the real trajectory, the canonically-normalized inflaton fluctuation is therefore defined by
\begin{equation}
T+\bar T
=
\exp\left(\frac{\phi}{M_P}\right) \, ,
\label{eq:ANR-T-canonical}
\end{equation}
in agreement with the general relation
$T+\bar T=\exp[\sqrt{2/(3\alpha)}\,\phi/M_P]$ evaluated at
$\alpha=2/3$.

\subsection{Matter Fields and Flipped SU(5)}
\label{sec:FSU5}
The model \cite{ANR1,ANR2} is a string-derived version of a flipped SU(5) $\times$ U(1) grand unified theory (GUT)~\cite{Barr,DKN,flipped2,AEHN}. 
This GUT contains three generations of SM
matter fields, each with a right-handed neutrino, which are each placed in
$\mathbf{10}$ ($F$), $\bar{\mathbf{5}}$ ($\bar f$), and $\mathbf{1}$ ($\sigma$) representations of SU(5).
The SU(5)$\times$U(1) GUT group is
broken to the SM via
$\mathbf{10}+\overline{\mathbf{10}}$ Higgs representations of SU(5) ($H+{\bar H})$. SM symmetry breaking proceeds via electroweak doublets contained
in $\mathbf{5}+\bar{\mathbf{5}}$ ($h+{\bar h}$) representations.
The superpotential is given by
\begin{align} \notag
W &=  \lambda_1^{ij} F_iF_jh + \lambda_2^{ij} F_i\bar{f}_j\bar{h} +
 \lambda_3^{ij}\bar{f}_i\ell^c_j h + \lambda_4 HHh + \lambda_5
 \bar{H}\bar{H}\bar{h}\\ 
&\quad   + \lambda_6^{ia} F_i\bar{H}\sigma_a + \lambda_7^a h\bar{h}\sigma_a
 + \lambda_8^{abc}\sigma_a\sigma_b\sigma_c + \mu^{ab}\sigma_a\sigma_b\,, 
\label{Wgen2} 
\end{align}
where the indices $i,j$ run over the three fermion families, and we have suppressed gauge group indices.
The first 3 terms of the superpotential (\ref{Wgen2}) correspond to the SM Yukawa couplings. 
The fourth and fifth terms account for the doublet-triplet separation.
The sixth term is related to the neutrino mass matrix. The seventh term is equivalent to the MSSM $\mu$-term. The last two terms determine the inflationary potential and are related to the superpotential (\ref{eq:ANR-W-yz}). These terms also enter into the neutrino mass matrix. The cosmology and phenomenology of this model has been explored in some detail \cite{EGNNO2,EGNNO3,EGNNO4,EGNNO5,EGNNO6, building}. While we will not review the details of the model here, we will pick out the components necessary for reheating. 

In the previous Section, we saw that the dominant contribution to reheating (when $\alpha \ne 1$ and/or the modular weights were not all equal to 1) was the trilinear superpotential term providing the top quark Yukawa coupling. This term is contained in the $\lambda_2$ coupling in (\ref{Wgen2}) (recall that ${\bar u}$ is contained in the $\bar{\mathbf{5}}$ representation of {\em flipped} SU(5) as opposed to the $\mathbf{10}$ in conventional SU(5)). In addition to this coupling, we will also be interested in the $\lambda_6$ coupling, which contains a direct superpotential coupling of the inflaton (we associate $y$ and $z$ with the singlets $\sigma_a$) to $\bar H$ and $\nu^c$ (contained in $F$). Thus, we will focus here on a limited part of the visible sector, 
namely that given by 
\begin{equation}
W_{\rm vis}
=
\lambda_2^{33} {\bar h} Q_3\bar u_3 + \lambda_6 \bar H \nu^c y \, .
\label{eq:ANR-top-Yukawa}
\end{equation}
In the string construction \cite{ANR1}, $\bar h$ is taken to be a twisted field 
from the same sector as $y$, whereas $Q_3$
$\bar u_3$, $\nu^c$, and ${\bar H}$ are twisted fields whose modular weights are associated with
the untwisted $z$ sector. The relevant K\"ahler potential may therefore be written
as
\begin{align}
K_{\rm string}
={}&
-\ln(S+\bar S)
-2\ln\left(
1-|y|^2-|{\bar h}|^2
\right)
\nonumber\\
&-
2\ln\left(
1-\frac{|z|^2}{2}
\right)
+
\frac{|Q_3|^2}
{\left(1-\frac{|z|^2}{2}\right)}
+
\frac{|\bar u_3|^2}
{\left(1-\frac{|z|^2}{2}\right)}
+
\frac{|\nu^c|^2}
{\left(1-\frac{|z|^2}{2}\right)}
+
\frac{|{\bar H}|^2}
{\left(1-\frac{|z|^2}{2}\right)} \, .
\label{eq:ANR-K-visible-yz}
\end{align}
The modular weights $n_{Q_3}=n_{\bar u_3}=n_{\nu^c}=n_{\bar H}=1$ appearing in
Eq.~\eqref{eq:ANR-K-visible-yz} characterize the dependence of these fields on the $z$-sector K\"ahler factor. They are not modular weights with respect to the inflaton modulus. When the dilaton and $z$ are stabilized at $z=0$, $S=\langle S\rangle$, the corresponding quadratic matter metrics are $Z_{Q_3}=Z_{\bar u_3} = Z_{\nu^c}=Z_{\bar H} =1$ and hence contain no dependence on the inflaton field $y$.

By contrast, expanding the $y$-sector logarithm to quadratic order in
${\bar h}$ gives
\begin{equation}
K_{\rm string}
\supset
\frac{2|\bar h|^2}{1-|y|^2}
+\cdots ,
\end{equation}
so that
\begin{equation}
Z_{\bar h}
=
\frac{2}{1-|y|^2} \, .
\label{eq:ANR-Higgs-metric-y}
\end{equation}
Although $Z_{H_u}$ depends on the inflaton background, this dependence
begins at quadratic order in $y$. In particular,
\begin{equation}
\left.
\frac{\partial Z_{\bar h}}
{\partial\,{\rm Re}\,y}
\right|_{y=0}
=0 \, .
\label{eq:ANR-Higgs-linear-zero}
\end{equation}
The same conclusion follows from the canonically-normalized Yukawa coupling. The top Yukawa interaction in Eq.~\eqref{eq:ANR-top-Yukawa} therefore does not generate a linear decay
of the inflaton about the vacuum.
We conclude that, for the specific string-motivated matter assignment in Eqs.~\eqref{eq:ANR-top-Yukawa} and \eqref{eq:ANR-K-visible-yz}, the top Yukawa coupling does not provide a linear perturbative decay channel for the $T$ inflaton. This conclusion is independent of whether it is derived in the symmetric $y$ basis.

Reheating in this realization must consequently arise from other
interactions, such as bilinear visible-sector terms, a modulus-dependent gauge kinetic function, anomaly-induced gauge and gaugino couplings, or additional superpotential operators whose transformed physical coefficients have nonvanishing first derivatives at the vacuum.

\subsection{Matter fields and modulus-dependent superpotential couplings}
\label{sec:matfields}

To compare this result with that from the analysis in the previous Section, it is useful
to transform from the symmetric basis to the $(T,C)$ basis.  We use Eq.~(\ref{eq:Tphi-to-y}),  so that 
\begin{equation}
    {\bar h}^{(y)} \; = \; \frac{{\bar h}^{(T)}}{T+\frac12} \, ,
\end{equation}
The $y$-sector K\"ahler potential then becomes
\begin{align}
-2\ln\left(
1-|y|^2-|{\bar h}^{(y)}|^2
\right)
={}&
-2\ln\left(
T+\bar T-|{\bar h}^{(T)}|^2
\right) +
2\ln\left|T+\frac12\right|^2,
\end{align}
up to a constant. The last term is removed by the
K\"ahler transformation
\begin{equation}
J(T)
=
-2\ln\left(T+\frac12\right),
\end{equation}
under which
\begin{equation}
W^{(T)}
=
e^{-J(T)}W^{(y)}
=
\left(T+\frac12\right)^2W^{(y)}.
\label{eq:ANR-W-Kahler-transformation}
\end{equation}
The top Yukawa term therefore transforms as
\begin{align}
\left(T+\frac12\right)^2
h_tH_u^{(y)}Q_3\bar u_3
&=
h_t
\left(T+\frac12\right)
H_u^{(T)}Q_3\bar u_3 \, .
\label{eq:ANR-top-W-T}
\end{align}
Thus, up to constant field normalizations, the top Yukawa operator in the
half-plane basis ($T,C$) is
\begin{equation}
W_{\rm top}^{(T)}
=
h_t
\left(T+\frac12\right)
H_u^{(T)}Q_3\bar u_3 \, .
\label{eq:ANR-top-Yukawa-T}
\end{equation}

More generally, one may consider a visible sector superpotential operator of the form
\begin{equation}
W_{\rm vis}^{(T)}
=
\left(T+\frac12\right)^\delta
\widehat W(\Phi_I) \, ,
\label{eq:general-delta-W}
\end{equation}
where $\widehat W$ is independent of $T$. If the matter metrics scale
as
\begin{equation}
Z_I\propto(T+\bar T)^{-n_I} \, ,
\end{equation}
then the canonically-normalized trilinear coupling scales as
\begin{equation}
\widehat W_{IJK}^{\rm phys}
\propto
e^{K/2}
\frac{
\left(T+\frac12\right)^\delta W_{IJK}
}{
\sqrt{Z_IZ_JZ_K}
} \, .
\end{equation}
Expanding along the real inflaton direction about
$T=\bar T=\tfrac12$, the coefficient of the linear inflaton
interaction is proportional to $n_I+n_J+n_K-3\alpha+\delta$. 
The decay width into the corresponding fermion--fermion--scalar final states is therefore obtained from the result for a $T$-independent superpotential by the replacement
\begin{equation}
n_I+n_J+n_K-3\alpha
\longrightarrow
n_I+n_J+n_K-3\alpha+\delta \, .
\label{eq:delta-shift-rule}
\end{equation}
in Eq.~(\ref{eq:Gamma-3body-fermion}).

The same shift appears in the associated four-scalar interaction. For a superpotential derivative carrying the common factor
$(T+\tfrac12)^\delta$, the four-body width becomes
\begin{equation}
\Gamma\bigl(
\delta T\rightarrow
\Phi_I\Phi_J\bar\Phi^{K}\bar\Phi^{M}
\bigr)
=
\frac{
\left(
n_I+n_J+n_L-3\alpha+\delta
\right)^2
}{\alpha}
\frac{
\left|
W^{IJL}\bar W_{LKM}
\right|^2m^3
}{
72(8\pi)^5M_P^2
}.
\label{eq:Gamma-4body-scalar-delta}
\end{equation}
Here it is understood that the holomorphic and antiholomorphic
superpotential derivatives originate from the same modulus-dependent operator, so that the linear variation of its absolute square produces
the single shift $+\delta$ in
Eq.~\eqref{eq:Gamma-4body-scalar-delta}.

Thus, despite the fact that the effective curvature is given by $\alpha = 2/3$, the superpotential defined in the basis defined by the K\"ahler potential in Eq.~(\ref{eq:ANR-K-visible-yz}) provides for a decay channel to Higgs and tops. 
The no-scale cancellation is most easily seen in the $T$ basis where ${\bar h}$ has K\"ahler modular weight 1, and the superpotential also carries modular weight 1
giving the cancellation precisely when $\alpha = 2/3$. In addition, if the gauge kinetic function is minimal in the $y$ basis, there is no decay to gauge bosons through the anomaly, as we discuss in more detail in the next Section. Nevertheless the inflaton does decay through its coupling to the neutrino sector. 

Inflaton decay in the flipped SU(5) model has been discussed in considerable detail \cite{EGNNO2,EGNNO3,EGNNO4,EGNNO5}.
GUT symmetry breaking occurs along a flat direction in which a combination, $\Phi$, of the  singlet components of $H$ and ${\bar H}$ obtains a vacuum expectation value. 
It was shown that reheating is completed in the SU(5) symmetric phase if the appropriate component of $|\lambda_{6}| \gtrsim
{\cal O}(10^{-4})$.
When $\langle \Phi\rangle < m$, the inflaton decays to $F$ and $\bar{H}$, with a rate given by
\beq\label{eq:decvan}
\Gamma(\delta T\rightarrow F_i\bar{H}) \;\simeq\; 10\times \frac{|\lambda_6^{i}|^2}{8\pi} \left( 1- \frac{\langle \Phi\rangle^2}{m^2}\right) m\,.
\eeq
The reheating
temperature in this case is given by
\begin{equation}
 T_{\rm RH} \simeq 1.7  \times 10^{15} ~\mathrm{GeV}
\times \lambda_6 \, ,
\label{eq:treh}
\end{equation}
indicating a direct relation between reheating and neutrino masses through 
$T_{\rm RH}$ and $\lambda_6$.
The importance of this $\lambda_6$ coupling \cite{EGNNO4} is amplified, now knowing that Yukawa couplings do not lead to reheating. This conclusion will be further strengthened when we see in the next Section that the coupling to gauge bosons is also absent.

\section{Decays to gauge bosons and gauginos}
\label{sec:T-gauge}

As we have seen above, in the $R+R^2$ model, even if the Higgs is conformally coupled to curvature, the inflaton may still decay through its coupling to the trace anomaly \cite{Gorbunov:2012ns}
\begin{align}
	\mathcal{L} \; = \; \frac{\phi}{\sqrt{6}\,M_P}
{T^\mu}_\mu. \, ,
 \label{OT}
\end{align}
In supergravity, this would correspond to a coupling of the K\"ahler potential to the trace anomaly \cite{Endo:2007sz,anom}
\begin{align}
	\mathcal{L} = -\frac{K}{6} {T^\mu}_\mu \, .
 \label{KT}
\end{align}
When the inflaton is identified with the volume modulus $T$, its coupling to the gauge sector receives two contributions in any given K\"ahler frame: a direct contribution from explicit inflaton dependence of the holomorphic Wilsonian gauge kinetic function $f_{\alpha\beta}$, and a one-loop contribution from the super-Weyl/K\"ahler and sigma model anomalies of supergravity. The separation between these contributions is frame-dependent, whereas their sum, and hence the physical inflaton-gauge amplitude, is not. Below we discuss separately the two contributions and demonstrate that the complete coupling is independent of the K\"ahler frame.

The relevant supergravity terms in the gauge sector are
\cite{Cremmer:1982en,egno4}
\begin{align}
{\cal L}_{G}
&=
-\frac{1}{4}
\bigl({\rm Re}\,f_{\alpha\beta}\bigr)
F^{\alpha}_{\mu\nu}F^{\beta\,\mu\nu}
+
\frac{i}{4}
\bigl({\rm Im}\,f_{\alpha\beta}\bigr)
F^{\alpha}_{\mu\nu}\widetilde F^{\beta\,\mu\nu}
\nonumber\\
&\quad
+
\left[
\frac{1}{4}\,e^{G/2}
(\bar f_{\alpha\beta})_{,J}
(G^{-1})^{J}{}_{K}G^{K}\,
\bar\lambda^{\alpha}_{L}\lambda^{\beta}_{R}
+\text{h.c.}
\right],
\label{eq:LG}
\end{align}
where $\widetilde F^{\mu\nu}=(i/2)\epsilon^{\mu\nu\rho\sigma}F_{\rho\sigma}$. If the gauge kinetic function depends on the volume modulus, $f(T)$, the first two terms determine the couplings of the scalar and pseudoscalar inflaton components to gauge bosons, while the last term gives the corresponding couplings to gauginos. Their contributions to the inflaton decay width are discussed further below. 

\subsection{K\"ahler frame invariance of the inflaton-gauge coupling}
For a product gauge group $G=\prod_a G_a$, and neglecting possible kinetic mixing between Abelian factors, the gauge kinetic function is block-diagonal. Within each simple factor $G_a$, gauge invariance restricts it to the form
\begin{equation}
f_{\alpha\beta}(\Phi)
=
f_a(\Phi)\,\delta_{\alpha\beta} \, ,
\end{equation}
where $\alpha,\beta$ are adjoint indices of $G_a$, and $\Phi$ denotes the neutral inflaton chiral multiplet. The gauge kinetic function $f_a(\Phi)$ is holomorphic. Its real part determines the inverse gauge coupling, while its imaginary part determines the corresponding $F\widetilde F$ interaction.

We denote the gauge-charged matter multiplets by $\Psi^A$, so that
\begin{equation}
K=K(\Phi,\bar\Phi,\Psi,\bar\Psi) \, .
\end{equation}
On the inflaton background, the charged matter fields are set to zero, and we define
\begin{equation}
\widehat K(\Phi,\bar\Phi)
\equiv
K(\Phi,\bar\Phi,\Psi=0,\bar\Psi=0) \, .
\end{equation}
For charged multiplets transforming in a representation $r$ of $G_a$, the corresponding matter K\"ahler metric is
\begin{equation}
Z^{(r)}_{A\bar B}(\Phi,\bar\Phi)
\equiv
\left.
\frac{\partial^2K}
{\partial\Psi^A\,\partial\bar\Psi^{\bar B}}
\right|_{\Psi=\bar\Psi=0},
\; \;
\text{where} \; A,\bar B\ \text{label multiplets in }r \, .
\label{eq:Zr-definition}
\end{equation}
Thus $Z^{(r)}$ is obtained by differentiating with respect to the charged matter fields, but after evaluation at the matter vacuum it depends on the inflaton background $(\Phi,\bar\Phi)$. We take the determinant over multiplet, or flavor, space. The trace over gauge indices is contained in the Dynkin coefficient $T_a(r)$. In particular, if
$Z^{(r)}_{A\bar B}=Z_r\delta_{A\bar B}$ for $N_r$ copies of $r$, then
\begin{equation}
\ln\det Z^{(r)}=N_r\ln Z_r \, .
\end{equation}
Invariance of the Lagrangian under the K\"ahler transformation will require the inclusion of one-loop contributions to the gauge kinetic function. 
Following~\cite{Kaplunovsky:1994fg}, 
we can define the 
coefficient of $-\frac14 F^a_{\mu\nu} F^{a~\mu\nu}$ as ${\cal F}_a$ including the one-loop contributions needed for K\"ahler transformations. 
The terms relevant for the field dependence and K\"ahler frame transformation of the one-loop physical inverse gauge coupling may be organized as
\begin{equation}
{\cal F}_a(\Phi,\bar\Phi)
=
{\rm Re}\,f_a(\Phi)
+
\frac{c_a}{16\pi^2}
\widehat K(\Phi,\bar\Phi)
-
\sum_r
\frac{T_a(r)}{8\pi^2}
\ln\det Z^{(r)}(\Phi,\bar\Phi)
+\cdots \, ,
\label{eq:Fa-field-dependent}
\end{equation}
where
\begin{equation}
c_a=T_R-T_a(G) \, ,
\qquad
T_R\equiv\sum_r N_rT_a(r) \, .
\label{catr}
\end{equation}
Here $T_a(G)$ is the Dynkin index of the adjoint representation and
$T_a(r)$ is that of the matter representation $r$.

Equation~\eqref{eq:Fa-field-dependent} displays only the terms needed for the discussion of the K\"ahler frame transformation. The ellipsis includes the ordinary momentum-dependent running and gauge field self-renormalization. These terms are separately invariant under the transformations considered below at the order relevant to the proof, although
they may carry additional model-dependent inflaton dependence. Unlike the holomorphic function $f_a(\Phi)$, the physical quantity ${\cal F}_a(\Phi,\bar\Phi)$ is real and generally nonholomorphic. Its variation along the physical inflaton direction determines the complete CP-even coupling to $F^a_{\mu\nu}F^{a\,\mu\nu}$.

Let us consider a combined transformation
\begin{align}
K
&\longrightarrow
K+J+\bar J \, ,
\nonumber\\
\Psi^{(r)}
&\longrightarrow
\Psi^{\prime(r)}
=
C_r(\Phi)\,\Psi^{(r)} \, ,
\label{eq:Kahler-matter-transformation}
\end{align}
where $J$ and $C_r$ are holomorphic. The matter metric transforms as
\begin{equation}
Z^{\prime(r)}
=
\bigl(C_r^{\dagger}\bigr)^{-1}
Z^{(r)}
C_r^{-1} \, ,
\label{eq:Z-transformation}
\end{equation}
and hence
\begin{equation}
\ln\det Z^{\prime(r)}
=
\ln\det Z^{(r)}
-
\ln\det C_r
-
\ln\det C_r^{\dagger} \, .
\label{eq:logdetZ-transformation}
\end{equation}
It was shown in~\cite{Kaplunovsky:1994fg} that the holomorphic gauge kinetic function transforms anomalously:
\begin{equation}
f_a^{\,\prime}
=
f_a
-
\frac{c_a}{8\pi^2}\,
J
-
\sum_r
\frac{T_a(r)}{4\pi^2}\,
\ln\det C_r \, .
\label{eq:KL-fW}
\end{equation}
Taking the real part of $f_a^{\,\prime}$,
and using Eqs.~\eqref{eq:Kahler-matter-transformation} and \eqref{eq:logdetZ-transformation}, we see that the three field-dependent terms in Eq.~\eqref{eq:Fa-field-dependent} transform as
\begin{align}
{\rm Re}\,f_a
&\longrightarrow
{\rm Re}\,f_a
-
\frac{c_a}{16\pi^2}(J+\bar J)
-
\sum_r
\frac{T_a(r)}{8\pi^2}
\ln\bigl|\det C_r\bigr|^2 \, ,
\nonumber\\[3pt]
\frac{c_a}{16\pi^2}\widehat K
&\longrightarrow
\frac{c_a}{16\pi^2}\widehat K
+
\frac{c_a}{16\pi^2}(J+\bar J) \, ,
\nonumber\\[3pt]
-\sum_r
\frac{T_a(r)}{8\pi^2}
\ln\det Z^{(r)}
&\longrightarrow
-\sum_r
\frac{T_a(r)}{8\pi^2}
\ln\det Z^{(r)}
+
\sum_r
\frac{T_a(r)}{8\pi^2}
\ln\bigl|\det C_r\bigr|^2 \, .
\end{align}
The terms proportional to $J+\bar J$ cancel between the gauge kinetic function and
K\"ahler contributions, while the terms involving $C_r$ cancel between the gauge kinetic function and matter wave-function contributions. Consequently,
\begin{equation}
{\cal F}_a'(\Phi,\bar\Phi)
=
{\cal F}_a(\Phi,\bar\Phi) \, .
\label{eq:Fa-invariant}
\end{equation}
The displayed field-dependent part of the physical inverse gauge coupling is therefore invariant under the combined K\"ahler transformation and holomorphic matter field redefinition. Its decomposition in Eq.~(\ref{eq:Fa-field-dependent}) into the tree-level gauge kinetic function, the K\"ahler anomaly, and matter wave-function contributions is frame-dependent, but their sum is not.

We now consider the contributions to inflaton decays of the various terms in $\cal{F}$. 

\subsection{Direct contribution from $f_{\alpha\beta}(T)$}
\label{sec:T-gauge-direct}
For a diagonal gauge kinetic function with explicit modulus dependence,
\begin{equation}
f_{\alpha\beta}(T)
=
f(T)\,\delta_{\alpha\beta},
\qquad
f(T)
=
\langle f\rangle
+
\left.
\frac{\partial f}{\partial T}
\right|_0
\bigl(T-\langle T\rangle\bigr)
+\cdots \, ,
\end{equation}
the strength of the direct inflaton-gauge coupling is characterized by
\cite{Endo:2006xg,egno4}
\begin{equation}
d_{g,T}
\equiv
\langle{\rm Re}\,f\rangle^{-1}
\left|
\left.
\frac{\partial f}{\partial T}
\right|_0
\right| \, ,
\label{eq:dgT-def}
\end{equation}
which vanishes when $f(T)$ is independent of $T$ in the K\"ahler frame under consideration.

Along the real inflaton trajectory, we use  the same canonically-normalized
modulus fluctuation $\delta T$ defined in Eq.~(\ref{eq:T-canonical-decay})
with the vacuum normalization $\langle T+\bar T\rangle=1$.
After canonical normalization of the gauge fields, the partial width into
gauge bosons belonging to the gauge-group factor $G_a$ is
\begin{equation}
\Gamma(\delta T\rightarrow A_aA_a)
=
\frac{d_{g,T}^{\,2}}{32\pi\alpha}
\left(\frac{N_a}{12}\right)
\frac{m^3}{M_P^2} \, ,
\label{eq:Gamma-direct}
\end{equation}
where $N_a=\dim G_a$ is the number of gauge bosons in the final-state
sector and $m$ is the inflaton mass. For $\alpha=1$, this reduces to
Eq.~(5.68) of Ref.~\cite{egno4}.

The direct decay to gauginos arising from the last term in $\mathcal{L}_G$ is proportional to $G^K$ and is therefore suppressed by $\mathcal{O}(m_{3/2}/m)$ at the amplitude level relative to the gauge boson channel, and by $\mathcal{O}[(m_{3/2}/m)^2]$ at the level of the
partial width \cite{Kallosh:2011qk}. This suppression is generic for the direct coupling in supergravity, but does not apply to the anomaly-induced gaugino channel discussed below.

\subsection{Anomaly-induced contribution}
\label{sec:T-gauge-anomaly}
We now extract the anomaly-induced decay directly from the field-dependent physical inverse gauge coupling introduced above. We first assume that the gauge kinetic function is independent of the inflaton:
\begin{equation}
\frac{\partial f_a}{\partial T}=0 \, .
\label{eq:fT-constant-anom}
\end{equation}
The inflaton dependence of ${\cal F}_a$ then arises entirely from the K\"ahler and matter wave-function terms,
\begin{equation}
{\cal F}_a^{\rm anom}
=
\frac{c_a}{16\pi^2}
\widehat K
-
\sum_r
\frac{T_a(r)}{8\pi^2}
\ln\det Z^{(r)} \, ,
\qquad
c_a=T_R-T_a(G) \, .
\label{eq:Fa-anom}
\end{equation}
Considering the no-scale K\"ahler potential~(\ref{eq:K-alpha-decays}), we set all the visible sector fields to zero after differentiation. The matter metrics take the form
\begin{equation}
Z_r
=
\kappa_r\,(T+\bar{T})^{-n_r} \, ,
\label{eq:Zr-noscale}
\end{equation}
where
\begin{equation}
n_r=
\begin{cases}
1, & \text{for untwisted matter},\\
n_a, & \text{for twisted matter},
\end{cases}
\end{equation}
and $\kappa_r$ is independent of $T$. In particular,
$\kappa_r=\alpha$ for untwisted fields, whereas
$\kappa_r=1$ for the twisted fields in Eq.~\eqref{eq:K-alpha-decays}. If there are
$N_r$ copies of the representation $r$, then
\begin{equation}
\ln\det Z^{(r)}
=
N_r\ln\kappa_r
-
N_rn_r\ln (T + \bar{T}) \, .
\end{equation}
The constant factor $\kappa_r$ therefore drops out of the inflaton
derivative.
Using the definition for $T_R$ in Eq.~(\ref{catr}),
we obtain
\begin{align}
\frac{\partial{\cal F}_a^{\rm anom}}{\partial T}
&=
-\frac{3\alpha c_a}{16\pi^2(T+ \bar{T})}
+
\frac{1}{T+\bar{T}}\sum_r
\frac{N_rn_rT_a(r)}{8\pi^2}
\nonumber\\
&=
\frac{1}{16\pi^2(T+ \bar{T})}
\left[
3\alpha\bigl(T_a(G)-T_R\bigr)
+
2\sum_rN_rn_rT_a(r)
\right] \equiv \frac{B_a^{(\alpha)}}{16\pi^2(T+\bar T)} \, .
\label{eq:dFa-anom-dT}
\end{align}
Along the real trajectory,
\begin{equation}
\left.
\frac{\partial T}{\partial\delta T}
\right|_0
=
\left.
\frac{\partial\bar T}{\partial\delta T}
\right|_0
=
\frac{\langle T+\bar T\rangle}
{\sqrt{6\alpha}} \, .
\end{equation}
Since the complete physical inverse gauge coupling ${\cal F}_a(T,\bar T)$ is generally nonholomorphic, the relevant variation is taken along the real modulus direction. We therefore generalize the tree-level coefficient $d_{g,T}$ to
\begin{equation}
d_{a,T}^{\rm eff}
\equiv
\left\langle{\cal F}_a\right\rangle^{-1}
\left|
\left.
\left(
\frac{\partial{\cal F}_a}{\partial T}
+
\frac{\partial{\cal F}_a}{\partial\bar T}
\right)
\right|_0
\right| \, .
\label{eq:d-eff-definition}
\end{equation}
For the purely holomorphic tree-level contribution, ${\cal F}_a={\rm Re}\,f_a$, this reduces along a CP-conserving real trajectory to
\begin{equation}
d_{a,T}^{\rm eff}
=
\left\langle{\rm Re}\,f_a\right\rangle^{-1}
\left|
\left.
\frac{\partial f_a}{\partial T}
\right|_0
\right|
=
d_{g,T} \, ,
\end{equation}
as in Eq.~(\ref{eq:dgT-def}).
After canonical normalization of the gauge fields, the interaction is
therefore 
\begin{equation}
{\cal L}_{\delta TFF}
=
-\frac{1}{4}
\frac{d_{a,T}^{\rm eff}}{\sqrt{6\alpha}}
\delta T\,
F^a_{\mu\nu}F^{a\,\mu\nu} \, ,
\label{eq:L-eff-from-F}
\end{equation}
up to the phase of the coupling. The corresponding partial width is the same
as for the direct contribution:
\begin{equation}
\Gamma(\delta T\to A_aA_a)
=
\frac{|d_{a,T}^{\rm eff}|^2}{32\pi\alpha}
\left(\frac{N_a}{12}\right)
\frac{m^3}{M_P^2} \, .
\label{eq:Gamma-gauge-master}
\end{equation}
To one-loop accuracy, $\left\langle{\cal F}_a\right\rangle^{-1} = g_a^2$, so that using Eq.~(\ref{eq:d-eff-definition})
\begin{equation}
d_{a,T}^{\rm anom}
=
\frac{g_a^2}{8\pi^2}
\left|B_a^{(\alpha)}\right| \; = \; \frac{\alpha_a}{2\pi} \left|B_a^{(\alpha)}\right| \, ,
\label{eq:d-anom-final}
\end{equation}
where $\alpha_a = g_a^2/4\pi$ is the generalized fine structure constant. 
Substitution into Eq.~\eqref{eq:Gamma-gauge-master} gives
\begin{equation}
\Gamma(\delta T\to A_aA_a)\big|_{\rm anom}
=
\frac{N_a g_a^4}{24576\pi^5\alpha}
\left|B_a^{(\alpha)}\right|^2
\frac{m^3}{M_P^2} \, .
\label{eq:Gamma-anom}
\end{equation}
The relation to the conventional trace anomaly is especially transparent for
the standard no-scale case, $\alpha=1$. Writing the beta function as
\begin{equation}
b_a
\equiv
3T_a(G)-T_R \, ,
\end{equation}
the anomaly coefficient becomes
\begin{align}
B_a^{(1)}
&=
3\bigl[T_a(G)-T_R\bigr]
+
2\sum_rN_rn_rT_a(r)=
b_a
+
2\sum_rN_r\bigl(n_r-1\bigr)T_a(r) \, .
\label{eq:B-alpha1-trace}
\end{align}
The first term is the usual one-loop supersymmetric beta function coefficient. It depends only on the charged particle content and gauge representations, and is independent of the modular weights. The second term is an additional contribution that measures how the modulus dependence of the charged-matter metrics differs from that of
untwisted no-scale matter. When all charged matter multiplets are untwisted, $n_r=1$, and in that case, $B_a^{(1)}=b_a$ and Eq.~(\ref{eq:Gamma-anom}) reduces to Eq.~(\ref{anomD}).
The reheating temperature in this case is
\begin{equation}
\trh
=
\left(
\frac{72}{5\pi^2g_{\rm RH}}
\right)^{1/4}
\left( \sum_a \frac{N_a b_a^2 \alpha_a^2}{1536 \pi^{3}}\right)^{1/2} \frac{m^{3/2}}{M_P^{1/2}} \simeq 4 \times 10^7~{\rm GeV} \, ,
\label{eq:Treh-Gamma2}
\end{equation}
where the numerical value is evaluated for MSSM inputs with $m = 3 \times 10^{13}$~GeV. For the MSSM, we use $b_i^{\rm MSSM} = \left(\frac{33}{5}, 1, -3 \right)$.

\subsection{Anomaly-induced coupling in the symmetric basis}
\label{sec:y-gauge-anomaly}
We now rewrite the anomaly-induced coupling in the symmetric SU(2,1)/SU(2)$\times$U(1) basis as in Eqs.~(\ref{eq:Tphi-to-y}) and (\ref{eq:y-to-Tphi}). In this subsection, we specialize to the no-scale case $\alpha=1$. It is convenient to define
\begin{equation}
D(y)
\equiv
1+\frac{y}{\sqrt{3}} \, ,
\label{eq:D-y2}
\end{equation}
and the volume modulus and the untwisted matter field are then related to the symmetric coordinates by
\begin{equation}
T+\bar T
=
\frac{1-|y|^2/3}{|D(y)|^2} \, ,
\qquad
\phi_i^{(T)}
=
D^{-1}\phi_i^{(y)} \, ,
\qquad
\phi_i^{(y)}\equiv y_i \, .
\label{eq:Y-y2}
\end{equation}
More generally, a charged matter multiplet of effective modular weight $n_r$ is rescaled according to
\begin{equation}
\Psi_r^{(y)}
=
D^{\,n_r}\Psi_r^{(T)} \, ,
\label{eq:Psi-y-transform}
\end{equation}
where $n_r=1$ for untwisted matter and $n_r=n_b$ for a twisted
multiplet of modular weight $n_b$. The transformation of the K\"ahler
potential may be written as
\begin{equation}
K^{(y)}
=
K^{(T)}+J+\bar J \, ,
\qquad
J(y)
=
-3\ln D(y) \, .
\label{eq:Kahler-y-transform}
\end{equation}
Correspondingly, the superpotential transforms as
\begin{equation}
W^{(y)}(y,y_i)
=
D(y)^{3}\,
W^{(T)}\bigl(T(y),\phi_i(y,y_i)\bigr) \, .
\label{eq:W-y-transform}
\end{equation}
Along the modulus-inflaton trajectory, $y_0=0$, and we define
\begin{equation}
S(y,\bar y)
\equiv
1-\frac{|y|^2}{3} \, .
\end{equation}
The matter metrics in the symmetric basis take the form
\begin{equation}
Z_r^{(y)}
=
\kappa_r\,S^{-n_r} \, ,
\label{eq:Z-y-symmetric}
\end{equation}
with $\kappa_r=\alpha$ for untwisted matter and $\kappa_r=1$ for the
twisted matter metrics considered above. For $N_r$ copies of representation
$r$,
\begin{equation}
\ln\det Z_r^{(y)}
=
N_r\ln\kappa_r
-
N_rn_r\ln S \, .
\end{equation}
The K\"ahler- and matter wave-function dependent part of the physical inverse gauge coupling is therefore
\begin{equation}
{\cal F}_{a,y}^{\rm anom}
=
\frac{c_a}{16\pi^2}\widehat K^{(y)}
-
\sum_r
\frac{T_a(r)}{8\pi^2}
\ln\det Z_r^{(y)}
=
\frac{B_a}{16\pi^2}\ln S
+\text{const.} \, ,
\label{eq:F-anom-y}
\end{equation}
where
$B_a$ is given by Eq.~(\ref{eq:dFa-anom-dT}) for $\alpha = 1$.
Since $S=1-|y|^2/3$, both the K\"ahler potential and the matter metrics
are even in $y$. Consequently,
\begin{equation}
\left.
\frac{\partial{\cal F}_{a,y}^{\rm anom}}{\partial y}
\right|_{y=y_i=0}
=
0 \, .
\label{eq:F-anom-y-zero}
\end{equation}
Thus the K\"ahler and matter wave-function terms do not generate a linear
inflaton coupling at the symmetric point.

The physical coupling nevertheless does not vanish. The combined K\"ahler
transformation and holomorphic matter rescaling are anomalous and induce a
field-dependent Wilsonian gauge kinetic function. Using the equations from the previous Section,
\begin{equation}
J=-3\ln D \, ,
\qquad
C_r=D^{n_r} \, ,
\end{equation}
gives
\begin{equation}
f_a^{(y)}(y)
=
f_a^{(T)}\bigl(T(y)\bigr)
-
\frac{B_a^{(\alpha)}}{8\pi^2}
\ln D(y) \, .
\label{eq:f-y-transformed}
\end{equation}
In particular, even when
\begin{equation}
\frac{\partial f_a^{(T)}}{\partial T}=0
\end{equation}
the symmetric basis function $f_a^{(y)}$ is not constant.

The equality of the physical inverse gauge couplings can be seen directly.
Using
\begin{equation}
T+\bar T
=
\frac{S}{|D|^2} \, ,
\end{equation}
the symmetric basis result may be written as
\begin{align}
{\cal F}_{a,T}
&=
{\rm Re}\,f_a^{(T)}
+
\frac{B_a^{(\alpha)}}{16\pi^2}
\ln(T+\bar T)
+\text{const.}
\nonumber\\
&=
{\rm Re}\,f_a^{(T)}
+
\frac{B_a^{(\alpha)}}{16\pi^2}\ln S
-
\frac{B_a^{(\alpha)}}{16\pi^2}\ln|D|^2
+\text{const.}
\nonumber\\
&=
{\rm Re}\,f_a^{(y)}
+
\frac{B_a^{(\alpha)}}{16\pi^2}\ln S
+\text{const.}
\nonumber\\
&=
{\cal F}_{a,y} \, .
\label{eq:F-basis-equality}
\end{align}
Thus the complete physical inverse gauge coupling is identical in the two
bases. In the half-plane basis the linear interaction is carried by the
K\"ahler and matter wave-function terms, whereas in the symmetric basis it is
carried entirely by the transformed Wilsonian gauge kinetic function.

At the symmetric vacuum,
\begin{equation}
\left.
\frac{\partial f_a^{(y)}}{\partial y}
\right|_0
=
-\frac{B_a^{(\alpha)}}{8\pi^2\sqrt{3}} \, .
\label{eq:dfdy2}
\end{equation}
The canonically-normalized real modulus is related to $y$ near the
vacuum by
\begin{equation}
\delta T
=
-\sqrt{2}\,{\rm Re}\,y+\cdots \, ,
\label{eq:y2-canonical}
\end{equation}
where the sign follows from Eq.~\eqref{eq:y-to-Tphi}. Hence
\begin{equation}
\left.
\frac{\partial{\cal F}_{a,y}}
{\partial\delta T}
\right|_0
=
\frac{B_a^{(\alpha)}}
{8\pi^2\sqrt{6}} \, ,
\label{eq:dFydeltaT}
\end{equation}
which is identical to the result obtained previously.

\section{$R^3$ contributions to inflaton decay}
\label{sec:R3}

The supergravity extension of a general $F(R)$ gravity was derived recently in \cite{Antoniadis:2026uzn}. It can be described as ordinary ${\cal N}=1$ supergravity coupled to three chiral superfields. Besides the scalaron of $F(R)$ gravity, $T$, and the goldstino (or `stabilizer') field $C$, there is a third superfield $X$ that decouples in the case of Starobinsky supergravity $R+\alpha_{S} R^2$,~\footnote{Here, $\alpha_S$ should not be confused with the parameter $\alpha$ appearing in the K\"ahler potential in Eq.~\eqref{eq:K-alpha-decays}.} and is an auxiliary field in the expansion around the vacuum $C=0$. To understand the physical meaning of the 3 superfields, $T$ is a Lagrange multiplier imposing the identification of $C$ with the chiral curvature superfield ${\cal R}$ while $X$ is the chiral projection of $\bar{\cal R}$. In the bosonic case, they are all related to one scalar field (the scalaron) linearizing the action, while in the supersymmetric case there are 2 independent fields because higher powers of $R$ emerge from non-chiral actions depending on ${\cal R}$ and $\bar{\cal R}$, unlike Einstein gravity.
It was shown that the point $C=0$ is stable for any $T$, and thus along the whole inflaton trajectory of $\text{Re}~T$, in the presence of the string dilaton, because $C$ acquires a large positive mass. Moreover, the absence of ghosts and tachyons in $F(R)$ supergravity implies two conditions on the first and third derivatives of $F(R)$ that generalize the conditions of the bosonic case and guarantee the stability of the scalar potential at ${\text{Im}}~T=C=0$, which is identical to that of  $F(R)$ gravity. It is important to note that the corresponding K\"ahler potential is of no-scale type for any $F(R)$, generalizing the case of Starobinsky supergravity.

We are interested here in extending the $F(R)$ supergravity in the presence of matter coupled to the metric of the Jordan frame, i.e., of $F(R)$ before expressing it as a 2-derivative action using Lagrange multipliers. For pedagogical reasons, we first derive the action in the case of $R+\alpha_S R^2$ supergravity, which can subsequently be extended using the same steps in a straightforward way to the more general case. We start in the absence of $R^2$ with an effective ${\cal N}=1$ supergravity of matter fields $\phi$ described by a K\"ahler potential $\hat{K}(\phi,\bar\phi)$ and a superpotential $\hat{W}(\phi)$ depending only on matter fields (gauge fields and interactions are not modified and are omitted for simplicity). Upon adding the supersymmetric $\alpha_S R^2$ term, the Lagrangian reads:
\begin{equation}
{\cal L}= -3\left[\left(e^{-\hat{K}/3}-6\alpha_S\frac{\cal R}{S_0}\frac{\bar{\cal R}}{\bar{S}_0}\right)S_0\bar{S}_0\right]_\text{D} 
+ \left(\left[\hat{W}S_0^3\right]_\text{F}+ \text{h.c.}\right)\,,
\label{eq:LR2}
\end{equation}
where $S_0$ is the chiral compensator and we use the notations and conventions of \cite{Antoniadis:2026uzn}. In particular, the subscripts D and F denote integrals over the full superspace (D-term) and chiral superspace (F-term), respectively, and the parameter $\alpha_S=1/(6m^2)$ with $m$ the inflaton mass, as denoted before. For $\alpha_S=0$, $\cal L$ reduces trivially to the initial supergravity of matter fields.

Introducing a Lagrange multiplier $T$ and identifying ${\cal R}/{S_0}=C/\sqrt{18\alpha_S}$, the Lagrangian \eqref{eq:LR2} can be rewritten as
\begin{align}
{\cal L}&= -3\left[\left(e^{-\hat{K}/3}-\frac{C{\bar{C}}}{3}\right)S_0\bar{S}_0\right]_\text{D} 
+ \left(\left[\left\{3T\left(\frac{C}{\sqrt{18\alpha_S}}-\frac{\cal R}{S_0}\right)+\hat{W}\right\}S_0^3\right]_\text{F}
+ \text{h.c.}\right)\nonumber\\
&=-3\left[\left(e^{-\hat{K}/3}+T+\bar{T}-\frac{C{\bar{C}}}{3}\right)S_0\bar{S}_0\right]_\text{D} 
+ \left(\left[\left\{\sqrt{3}mTC+\hat{W}\right\}S_0^3\right]_\text{F}+ \text{h.c.}\right)\,,
\label{eq:LR2TC}
\end{align}
where in the second line we used a supersymmetric identity. It follows that the initial supergravity defined by $\hat{K}$ and $\hat{W}$ is modified in the presence of $R^2$ to $K$ and $W$ that depend now also on $T$ and $C$:
\begin{align}
K=-3\ln\left(e^{-\hat{K}(\phi)/3}-1+T+\bar{T}-\frac{C{\bar{C}}}{3} \right)\,;\quad W=\sqrt{3}mC\left(T-\frac12\right)+\hat{W}(\phi)\,,
\label{eq:LR2KW}
\end{align}
where $T$ has been redefined by $T\to T-1/2$.
For matter with a no-scale K\"ahler potential $\hat{K}=-3\ln(1-\sum_\alpha |\phi_\alpha|^2/3)$, one obtains the expression given in Eq.~\eqref{nsK} for $K$. Indeed, one obtains the same form for any $\hat{K}$ when expanding about canonical kinetic terms.

The generalization to $F(R)$ is now straightforward, following the same steps as in \cite{Antoniadis:2026uzn}. As a concrete example,  we next consider the addition of an $R^3$ term.

\subsection{$R^3$ corrections}

We start from the supergravity embedding of $R + \alpha_S R^2 + \beta R^3$ in the formulation of~\cite{Antoniadis:2026uzn}, augmented by a matter superfield
$\phi_m$ coupled inside the no-scale logarithm, as was described above:
\begin{align}
K
&=
-3\ln\!\left[
T+\bar T
-\frac{\phi_m\bar\phi_m}{3}
-\frac{C\bar C}{3}
\left(
1+\frac{\beta}{m^2}(X+\bar X)
\right)
\right],
\label{eq:K-full}
\\
W
&=
\sqrt{3}mC
\left[
T-\frac12
+\frac{\beta}{8m^4}X^2
\right] \, ,
\label{eq:W-full}
\end{align}
where $X$ is the auxiliary chiral superfield encoding the $R^3$ deformation (non-propagating around $C=0$), and $\beta$ is the dimensionless coupling normalized as in~\cite{Gialamas:2025ofz, Antoniadis:2026uzn}.

It is important that the $X$-dependent factor in
Eq.~\eqref{eq:K-full} multiplies only the sgoldstino bilinear
$C\bar C/3$. In particular, the additional matter terms
$\phi_m\bar\phi_m/3$ are not dressed by $X$ or $\beta$.
Restricting to the real inflationary trajectory, $T_I=X_I=C=0$, the scalar potential before eliminating $X_R$ is~\footnote{The notation $T_R$ in this Section denotes the real component of the scalar field $T$ and should not be confused with the total matter Dynkin index $T_R\equiv\sum_r N_rT_a(r)$ introduced earlier.}
\begin{equation}
V(T_R,X_R)
 \; = \; \frac{3 X_R}{16m^4 (2T_R-\frac13|\phi_m|^2)^2} \left[4m^4(2T_R - 1) -m^2 X_R - \beta X_R^2 \right]\, .
\label{eq:V-TR-XR}
\end{equation}
Since $X_R$ has no kinetic term on this trajectory, its equation of motion is the algebraic stationarity condition
$\partial V/\partial X_R=0$. Besides the branch for which $D_CW=0$, the branch continuously connected to the $R+R^2$ theory satisfies
\begin{equation}
3\beta X_R^2
+2m^2X_R
-8m^4\left(T_R-\frac12\right)
=0 \, ,
\label{eq:XR-equation}
\end{equation}
and is independent of $\phi_m$.
Its solution is
\begin{equation}
X_R^-(T_R)
=
-\frac{m^2}{3\beta}
\left[
1-
\sqrt{
1+24\beta\left(T_R-\frac12\right)
}
\right] \, ,
\label{eq:XR-solution}
\end{equation}
where the branch has been chosen such that the limit
$\beta\rightarrow0$ is regular. For small $|\beta|$,
\begin{equation}
X_R^-(T_R)
=
4m^2\left(T_R-\frac12\right)
-24\beta m^2
\left(T_R-\frac12\right)^2
+\mathcal O(\beta^2) \, .
\label{eq:X-expansion}
\end{equation}
The function multiplying $C\bar C/3$ in the K\"ahler potential is
\begin{equation}
F_C(T,\bar T)
\equiv
1+\frac{\beta}{m^2}(X+\bar X) \, .
\end{equation}
Along the real trajectory,
$X+\bar X=2X_R^-(T_R)$ and
$T_R-\tfrac12=\tfrac12(T+\bar T-1)$, so that
\begin{align}
F_C(T,\bar T)
={}&
1+4\beta(T+\bar T-1)-12\beta^2(T+\bar T-1)^2
+\mathcal O(\beta^3) \, .
\label{eq:FC-expansion}
\end{align}
Consequently, to quadratic order in $C$ and $\phi_m$, the effective
K\"ahler potential along the real trajectory takes the form
\begin{align}
K_{\rm eff}
=
-3\ln\!\Big[
&T+\bar T-\frac{\phi_m\bar\phi_m}{3}-\frac{C\bar C}{3}
\big\{
1+4\beta(T+\bar T-1)
+\mathcal O(\beta^2)
\big\}
\Big] \, .
\label{eq:K-eff}
\end{align}
At $\beta=0$, this reduces to the standard Cecotti/no-scale form,
\begin{equation}
K
=
-3\ln\left(
T+\bar T-\frac{C\bar C}{3}-\frac{\phi_m\bar\phi_m}{3}
\right) \, .
\end{equation}
To leading order in $\beta$, the $R^3$-deformed no-scale supergravity is
captured by a modification of the $\alpha=1$ Cecotti model (see Eq.~(\ref{CSM})) in which the sgoldstino bilinear $C \bar{C}/3$ receives a K{\" a}hler-potential dressing factor
$\big[1 + 4\beta\,(T+\bar T - 1)\big]$, and the superpotential acquires the
holomorphic correction $2\beta\,(T - \tfrac{1}{2})^2$. Higher orders in
$\beta$ follow systematically from~\eqref{eq:X-expansion}.

\section{Summary and conclusions}
\label{sec:summ}
We have studied perturbative reheating in no-scale supergravity models in which the inflaton is identified with the real component of the volume modulus $T$. In the minimal no-scale realization with scalar manifold curvature $\mathcal R=2/3$, the leading inflaton couplings to conformally-embedded visible matter exhibit a well-known no-scale cancellation. In particular, for a $T$-independent trilinear superpotential coupling among three untwisted fields, the coefficient of the linear inflaton interaction vanishes.

We generalized the matter decay amplitudes to the K\"ahler potential with an $\alpha$ dependence, given by Eq.~(\ref{eq:K-alpha-decays}), whose scalar manifold curvature is $\mathcal R=2/(3\alpha)$. After canonical normalization, an untwisted field has effective $T$-modular weight one. For a trilinear visible operator with no explicit modulus dependence, the corresponding three-body decay amplitude is therefore controlled by a factor $n_I+n_J+n_K-3\alpha$ (see Eq.~(\ref{eq:Gamma-3body-fermion})). The conformal cancellation is recovered when the relevant sum of modular weights equals $3\alpha$. The dominant tree-level decay rate then scales as $m^3/M_P^2$, up to visible couplings and modular weights.

The coupling of the modulus to gauge bosons receives both a direct
contribution from the holomorphic gauge kinetic function and a
one-loop contribution from the super-Weyl/K\"ahler and sigma model
anomalies. Although the separation of these terms depends on the
K\"ahler frame and on holomorphic matter-field redefinitions, their sum is invariant. For a modulus-independent tree-level gauge kinetic function, the anomaly-induced channel therefore provides an
irreducible reheating mechanism with a minimal reheating temperature of $T_{\rm RH} = 2.3 \times 10^{8} \, \rm{GeV}$. 
However, this result does depend on which K\"ahler frame one choose to take a field-independent gauge kinetic function. For example, in the symmetric basis with a minimal gauge kinetic function, the coupling of the canonical inflaton to gauge fields vanishes at the tree level and there is no contribution from the anomalies.   

We used the decay rates (as a function of $\alpha$) to determine the reheating history
self-consistently. For every value of $\alpha$, the scalar amplitude fixes the normalization of the inflationary potential and hence the physical inflaton mass. The microscopic decay rate then fixes the radiation density, the averaged reheating equation of state, and the number of $e$-folds $N_*$. Consequently, $\alpha$ and $T_{\rm RH}$ are not independent phenomenological parameters in this realization. This produces a single self-consistent trajectory in the $(n_s,r)$ plane, as seen in Fig.~\ref{fig:nsrplot1}. At the conformal no-scale point $\alpha=1$, the tree-level trilinear matter channel vanishes and the reheating floor is set by the anomaly-induced coupling to the gauge sector. We find that the generalized no-scale model is compatible with {\it Planck} \cite{Planck} data for $\alpha \le 20$ and with the combination of SPT, {\it Planck}, ACT and BICEP/Keck data \cite{SPT-3G:2025bzu} for $1 \le \alpha \le 20$.

We also considered the string-derived ANR model~\cite{ANR1,ANR2,ANO}, for which
$\alpha_{\rm ANR}=2/3$. This example illustrates the importance of
transforming both the K\"ahler potential and the superpotential
consistently. A top Yukawa interaction that is constant in the symmetric
basis acquires an explicit factor $T+\tfrac12$ after a transformation to the ($T,C)$ basis. This induced modulus dependence cancels the apparent contribution from the matter metrics at linear order, so the ordinary top Yukawa
interaction does not provide for reheating despite having $\alpha \ne 1$. By contrast, the
$\lambda_6$ operator contains an explicit coupling of the inflaton to matter and has a nonzero first derivative at the vacuum, providing a direct connection between inflaton decay and the neutrino
sector.

Finally, we examined an $R^3$ deformation of Starobinsky
supergravity. The additional chiral field $X$ is auxiliary on the
inflationary trajectory and is eliminated through its algebraic
equation of motion. The resulting deformation changes the inflaton
potential and the sgoldstino metric \cite{Antoniadis:2026uzn}. However, when visible matter is introduced independently through
$-\phi_m\bar\phi_m/3$ inside the no-scale logarithm, its metric remains $Z_{\phi_m}\propto(T+\bar T)^{-1}$, and its effective modular weight remains unity. Thus the $R^3$
deformation does not by itself generate a new explicit
$\beta$-dependent decay vertex into such visible fields. Its effects on visible-sector reheating are indirect, through the modified inflationary normalization and background evolution, unless additional operators couple the visible sector directly to $X$ or to the deformation.

Our results demonstrate that reheating is a sensitive probe of the
K\"ahler geometry, the matter embedding, and the K\"ahler frame in which microscopic superpotential couplings are specified. A consistent treatment of these ingredients removes the apparent freedom to choose the reheating temperature independently of the inflationary model and allows CMB observables to test the underlying no-scale construction.

\section*{Acknowledgments}
We would like to thank M. Garcia and W. Ke for helpful conversations. I.A. and D.V.N. would like to thank the Center for Cosmology and Particle Physics at New York University, where part of this work was performed, for hospitality.
This research of I.A. was supported in part by the Higher Education and Science Committee of MESCS RA (Research Project N 24RL-1C036). The work of J.E. was supported by the United Kingdom STFC Grant ST/T000759/1. The work of K.A.O. was supported in part by DOE grant DE-SC0011842 at the University of Minnesota. The work of S.V. was supported by the Kavli Institute for Cosmological Physics at the University of Chicago.


\appendix

\section{Jordan frame interpretation for \texorpdfstring
{$\alpha\neq1$}{alpha neq 1}}
\label{app:jordanframe}
In the ordinary Starobinsky case, the relation between the Jordan frame $F(R)$ description and the no-scale supergravity description is well understood. Starting from 
\begin{equation}
{\cal A}
=
-\frac12\int d^4x\sqrt{-g}\left(R-F(R)\right) \, ,
\end{equation}
with
\begin{equation}
F(R)=\frac{R^2}{6m^2} \, ,
\end{equation}
one may introduce an auxiliary field $\Phi$ and rewrite the action as
\begin{equation}
{\cal A}
=
-\frac12\int d^4x\sqrt{-g}
\left[
R-F(-\Phi)-F'(-\Phi)(R+\Phi)
\right] \, .
\end{equation}
Eliminating $\Phi$ gives $\Phi=-R$, recovering the original theory. The
Einstein frame metric is obtained through the conformal transformation
\begin{equation}
\tilde g_{\mu\nu}=e^{2\Omega}g_{\mu\nu},
\qquad
e^{2\Omega}=1-F'(-\Phi) \, .
\end{equation}
For the Starobinsky choice this gives
\begin{equation}
e^{2\Omega}=1+2 \alpha_S \Phi,
\qquad
\alpha_S=\frac{1}{6m^2} \, ,
\end{equation}
and the canonically-normalized scalar is
\begin{equation}
\phi=\sqrt6\,M_P\,\Omega \, .
\label{canO}
\end{equation}
The Einstein frame potential is then
\begin{equation}
V(\phi)
=
\frac34 m^2 M_P^2
\left(1-e^{-\sqrt{2/3}\, \phi/M_P} \right)^2 \, .
\end{equation}
This is the usual Starobinsky potential and corresponds to the standard
$\alpha=1$ no-scale model.

One might try to generalize this construction by choosing an $F(R)$ function such that
\begin{equation}
1-F'(-\Phi)=(1+2\alpha_S\Phi)^\alpha \, ,
\end{equation}
or equivalently
\begin{equation}
e^{2\Omega}=(1+2\alpha_S\Phi)^\alpha \, .
\end{equation}
However, this does not reproduce the kinetic geometry of a no-scale
$\alpha$-attractor. Defining
\begin{equation}
Y\equiv 1+2\alpha_S\Phi \, ,
\end{equation}
the conformal factor implies
\begin{equation}
\Omega=\frac{\alpha}{2}\ln Y \, .
\end{equation}
Since the Einstein frame scalar from an $F(R)$ theory is universally given by Eq.~(\ref{canO}), 
one obtains
\begin{equation}
\phi
=
\alpha\sqrt{\frac32}\,M_P\ln Y \, .
\end{equation}
In contrast, in a no-scale $\alpha$-attractor with
\begin{equation}
K=-3\alpha\ln(T+\bar T) \, ,
\end{equation}
the canonically-normalized modulus is
\begin{equation}
\phi_{\rm ns}
=
\sqrt{\frac{3\alpha}{2}}\,
M_P\ln(T+\bar T) \, .
\end{equation}
Thus the kinetic coefficients agree only for
\begin{equation}
\alpha=1 \, .
\end{equation}
For $\alpha\neq1$, the naive $F(R)$ deformation gives a kinetic coefficient
proportional to $\alpha^2$, whereas the no-scale $\alpha$-attractor geometry requires a coefficient proportional to $\alpha$.

The same point can be seen from the K\"ahler potential. If
\begin{equation}
e^{2\Omega}=Y^\alpha \, ,
\end{equation}
then
\begin{equation}
\ln Y=\frac{2\Omega}{\alpha} \, .
\end{equation}
Therefore
\begin{equation}
K=-3 \, \alpha\ln Y=-6\Omega \, .
\end{equation}

However, we note that the corresponding Einstein frame
potential is not the usual $\alpha$-attractor potential. If one chooses
\begin{equation}
F(R)
=
R+\frac{1}{2\alpha_S(1+\alpha)}(1-2\alpha_SR)^{1+\alpha}
-\frac{1}{\alpha_S(1+\alpha)} \, ,
\end{equation}
so that the vacuum energy vanishes at $\Phi=0$, then
\begin{equation}
1-F'(-\Phi)=Y^\alpha ,
\qquad
Y=1+2\alpha_S\Phi \, ,
\end{equation}
and the Einstein frame potential is
\begin{equation}
V
=
\frac{M_P^2}{4\alpha_S}
\left[
\frac{\alpha}{1+\alpha}Y^{1-\alpha}
-
Y^{-\alpha}
+
\frac{1}{1+\alpha}Y^{-2\alpha}
\right] \, .
\end{equation}
Using $Y=e^{2\Omega/\alpha}$, this can be written as
\begin{equation}
V
=
\frac32m^2 M_P^2
\left[
\frac{\alpha}{1+\alpha}
e^{\frac{2(1-\alpha)}{\alpha}\Omega}
-
e^{-2\Omega}
+
\frac{1}{1+\alpha}e^{-4\Omega}
\right] \, .
\end{equation}
For $\alpha=1$, this reduces to
\begin{equation}
V
=
\frac34m^2 M_P^2(1-e^{-2\Omega})^2
=
\frac34m^2 M_P^2
\left(1-e^{-\sqrt{2/3}\,\phi/M_P}\right)^2 \, ,
\end{equation}
as expected. For $\alpha\neq1$, however, this potential is not the standard no-scale $\alpha$-attractor potential.

We therefore cannot identify a general Jordan frame non-minimal coupling parameter with the no-scale parameter $\alpha$. The robust correspondence occurs only at the special cancellation point:
\begin{equation}
\alpha=1
\qquad\Longleftrightarrow\qquad
\xi=\frac16 \, ,
\end{equation}
where the no-scale cancellation of tree-level matter decays is analogous to the conformal cancellation of scalaron decays into conformally-coupled scalars. Away from this point, the $\alpha$-deformed no-scale theory should be treated directly in the Einstein frame supergravity description rather than inferred from a naive $F(R)$ deformation.

\end{document}